\documentclass[fleqn,usenatbib]{mnras}

\usepackage{newtxtext}
\usepackage[varvw]{newtxmath}

\usepackage[T1]{fontenc}

\DeclareRobustCommand{\VAN}[3]{#2}
\let\VANthebibliography\thebibliography
\def\thebibliography{\DeclareRobustCommand{\VAN}[3]{##3}\VANthebibliography}

\usepackage{graphicx}	
\usepackage{amsmath}	
\usepackage{tablefootnote}

\def\Msun{\rm M_\odot}
\def\msun{{\Msun}}

\def\Msun{\rm M_\odot}

\def\HI{\hbox{H~$\scriptstyle\rm I\ $}}

\title[21cm-LAE cross-correlations]{On the general nature of 21cm-Lyman-$\alpha$ emitter cross-correlations during reionisation}

\author[Hutter et al.]{
Anne Hutter$^{1,2}$\thanks{E-mail: anne.hutter@nbi.ku.dk},
Caroline Heneka$^{3}$\thanks{E-mail: heneka@thphys.uni-heidelberg.de},
Pratika Dayal$^{4}$,
Stefan Gottl\"ober$^{5}$,
Andrei Mesinger$^{6}$,
\newauthor
Maxime Trebitsch$^{4}$,
Gustavo Yepes$^{7,8}$
\\
$^{{1}}$ Cosmic Dawn Center (DAWN) \\
$^{2}$ Niels Bohr Institute, University of Copenhagen, Jagtvej 128, DK-2200, Copenhagen N, Denmark \\
$^{3}$ Institute of Theoretical Physics (ITP), Heidelberg University, Philosophenweg 16, 69120 Heidelberg, Germany \\
$^{4}$ Kapteyn Astronomical Institute, University of Groningen, P.O. Box 800, 9700 AV Groningen, The Netherlands \\
$^{5}$ Leibniz-Institut f\"ur Astrophysik, An der Sternwarte 16, 14482 Potsdam, Germany \\
$^{6}$ Scuola Normale Superiore, Piazza dei Cavalieri 7, I-56126 Pisa, Italy \\
$^{7}$ Departamento de Fisica Teorica, Modulo 8, Facultad de Ciencias, Universidad Autonoma de Madrid, 28049 Madrid, Spain\\
$^{8}$ CIAFF, Facultad de Ciencias, Universidad Autonoma de Madrid, 28049 Madrid, Spain
}

\date{Accepted 2023 July 31. Received 2023 July 21; in original form 2023 June 5}

\pubyear{2023}

\begin{document}
\label{firstpage}
\pagerange{\pageref{firstpage}--\pageref{lastpage}}
\maketitle

\begin{abstract}
We explore how the characteristics of the cross-correlation functions between the 21cm emission from the spin-flip transition of neutral hydrogen (\HI) and early Lyman-$\alpha$ (Ly$\alpha$) radiation emitting galaxies (Ly$\alpha$ emitters, LAEs) depend on the reionisation history and topology and the simulated volume. For this purpose, we develop an analytic expression for the 21cm-LAE cross-correlation function and compare it to results derived from different {\sc astraeus} and {\sc 21cmfast} reionisation simulations covering a physically plausible range of scenarios where either low-mass ($\lesssim10^{9.5}\msun$) or massive ($\gtrsim10^{9.5}\msun$) galaxies drive reionisation. Our key findings are: (i) the negative small-scale ($\lesssim~2$~cMpc) cross-correlation amplitude scales with the intergalactic medium's (IGM) average \HI fraction ($\langle\chi_\mathrm{HI}\rangle$) and spin-temperature weighted overdensity in neutral regions ($\langle1+\delta\rangle_\mathrm{HI}$); (ii) the inversion point of the cross-correlation function traces the peak of the size distribution of ionised regions around LAEs; (iii) the cross-correlation amplitude at small scales is sensitive to the reionisation topology, with its anti-correlation or correlation decreasing the stronger the ionising emissivity of the underlying galaxy population is correlated to the cosmic web gas distribution (i.e. the more low-mass galaxies drive reionisation); (iv) the required simulation volume to not underpredict the 21cm-LAE anti-correlation amplitude when the cross-correlation is derived via the cross-power spectrum rises as the size of ionised regions and their variance increases. Our analytic expression can serve two purposes: to test whether simulation volumes are sufficiently large, and to act as a fitting function when cross-correlating future 21cm signal Square Kilometre Array and LAE galaxy observations.
\end{abstract}




\section{Introduction}

Our Universe underwent the last major phase transition during its first billion years when the ultraviolet (UV) photons from the first stars and galaxies ionised the neutral hydrogen (\HI) in the intergalactic medium (IGM). During this Epoch of Reionisation (EoR), ionised regions grew and merged around galaxies until the IGM was ionised by $z\simeq5.3$ \citep{Keating2020, Zhu2021, Bosman2021, Qin2021, Dayal2018}. However, the exact timing of the reionisation process and the topology of the ionised IGM, i.e. the evolution of the spatial distribution of ionised regions within the cosmic web structure, remain uncertain. Both go back to our limited knowledge about the properties of the first galaxies and whether the majority of \HI ionising photons emerged from the few massive galaxies in the densest regions or from the numerous low-mass galaxies that are more homogeneously distributed in the cosmic web structure.

In the past years, the rising number of observed high-redshift galaxies and precision measurements of the cosmic microwave background (CMB) have started to paint a picture wherein reionisation occurs has a midpoint around $z\simeq7-8$ \citep{Planck2020, Goto2021, Maity2022}. A robust tracer of the IGM ionisation state is the presence or absence of the \HI sensitive Lyman-$\alpha$ (Ly$\alpha$) emission line in detected galaxy spectra. The number density, fraction and spatial distribution of galaxies with observable Ly$\alpha$ emission, so-called Ly$\alpha$ emitters (LAEs), track the mean \HI fraction ($\langle\chi_\mathrm{HI}\rangle$) in the IGM \citep[e.g.][]{Mesinger2008, Dayal2011, Dijkstra2014, Hutter2014, Schenker2014, Pentericci2014, Pentericci2018, Fuller2020} and spatial distribution of ionised regions \citep[e.g.][]{Jensen2013, Mesinger2015, Hutter2017, Hutter2020, Qin2022, Castellano2016, Castellano2018}. However, the fraction of Ly$\alpha$ radiation transmitted through the IGM is sensitive to the shape of the Ly$\alpha$ line emerging from a galaxy, which again is subject to the gas density and velocity distribution of its interstellar and circumgalactic media \citep[e.g.][]{Verhamme2015, Dijkstra2016, Gronke2017, Kimm2019, Kakiichi2021}.

Fortunately, current and forthcoming radio interferometers, such as the Square Kilometre Array\footnote{Square Kilometre Array, \url{https://www.skatetelescope.com}} \citep[SKA;][]{Carilli2004}, Hydrogen Epoch of Reionisation Array \citep[HERA;][]{DeBoer2017}, Murchison Widefield Array\footnote{Murchison Widefield Array, \url{http://www.mwatelescope.org}} \citep[MWA;][]{Li2019, Barry2019} and Low Frequency Array\footnote{Low Frequency Array,  \url{http://www.lofar.org}} \citep[LOFAR;][]{Patil2017, Mertens2020}, will detect the cosmic \HI 21cm signal that traces the topology of the ionised regions in the IGM, with SKA being expected to have sufficient angular resolution and sensitivity to provide us with real-space 21cm maps. Statistical analyses applied to the 21cm signal measured in reciprocal space alone, such as the 21cm auto power spectra, will constrain our models of reionisation and the underlying galaxy population driving this phase transition. However, they rely on the accurate removal of various 21cm signal foregrounds interfering with the EoR signal \citep[e.g.][]{Shaver1999, Barry2016, Trott2016, Patil2016, Patil2017, Mertens2018, Mertens2020}. Theoretically, cross-correlating the 21cm signal with galaxy surveys eases the removal of bright 21cm foregrounds, as the only foregrounds that survive the cross-correlation are those arising from the cosmological volume of the galaxy survey,\footnote{For example, low-redshift interlopers in high-redshift galaxy surveys could correlate with point sources that are part of the 21cm foregrounds and generate false correlation signatures.} confirming the reality of the cosmological 21 cm signal \citep{Beane2019, Furlanetto_Lidz2007}. 
In practice, however, 21cm foregrounds will inflate the variance of 21cm-galaxy cross-correlations compared to a hypothetical 21cm foreground-free survey. For this reason, 21cm foreground mitigation is still desirable and its quality increases for larger survey areas \citep{Liu2020}.
Thus, efforts have concentrated on investigating the power of cross-correlations between the 21cm signal and galaxies \citep{Furlanetto_Lidz2007, Wyithe2007, Park2014}. 
Furthermore, as the Ly$\alpha$ line detected in spectroscopic or narrow-band surveys allows for more precise redshift estimates of the selected galaxies than broad-band Lyman break galaxy surveys, a strong focus has been on exploring the constraining power of 21cm-LAE cross-correlations, either in terms of cross-power spectra or cross-correlation functions \citep{Wiersma2013, Sobacchi2016, Vrbanec2016, Hutter2017, Heneka2017, Kubota2018, Hutter2018b, Heneka2020, Vrbanec2020, Weinberger2020}.
Indeed, for various reionisation scenarios and LAE models, 21cm-LAE cross-correlations exhibit $\langle\chi_\mathrm{HI}\rangle$-sensitive signatures, such as the cross-correlation or cross-power amplitude and the scale where the cross-power spectrum switches signs or the cross-correlation function changes its curvature. These signatures stem from the large-scale anti-correlation (correlation) between the 21cm signal in emission (absorption) and the LAEs located in ionised regions \citep[see e.g.][]{Heneka2020} as well as the corresponding cross-correlations tracing the size of ionised regions around LAEs. However, despite these fundamental relations, the values and signs for the small-scale cross-correlation function and power spectra differ among different works. While the change in sign reflects whether the 21cm signal is predominantly in absorption or emission, it remains unclear whether the remaining differences are signatures of different reionisation scenarios and LAE models or arise from limited simulated volumes or the chosen normalisations for the underlying 21cm and LAE number density fluctuations. Only a thorough understanding of the 21cm-LAE cross-correlations will allow us to tighten constraints on the reionisation history and topology as well as the nature of Ly$\alpha$ emitting galaxies and assess which supplementary statistics and/or data might be required further.

We address this question in this paper. For this purpose, we derive the small-scale analytic limit of the 21cm-LAE cross-correlation function and propose an analytic fitting function. We compare the analytic predictions with results from different simulations with {\sc astraeus} \citep{Hutter2022} and {\sc 21cmfast} \citep{Mesinger2016}, and analyse: What reionisation characteristics (e.g. ionisation history and topology) does the small-scale amplitude of the 21cm-LAE cross-correlation function trace? Which feature in the 21cm-LAE cross-correlation function tracks the typical size of the ionised regions? What are the effects of self-shielded regions around LAEs and limited simulation volumes?

This paper is organised as follows. In Section \ref{sec:21cm_LAE_cross_correlations}, we derive the analytic limits and model for the 21cm-LAE cross-correlation function during reionisation. We then compare the results from different {\sc astraeus} and {\sc 21cmfast} reionisation simulations to the analytic predictions derived in Section \ref{sec:21cm_LAE_cross_correlations} and assess the dependence of the 21cm-LAE cross-correlation function amplitude on the reionisation topology in Section \ref{sec:results_from_simulations}. In Section \ref{sec:impact_of_limited_volumes}, we investigate the effects of limited simulation volumes on the 21cm-LAE cross-correlation function and discuss the results from existing literature in Section \ref{sec:comparison_to_previous_work}. We conclude in Section \ref{sec:conclusions}.

\section{21cm-LAE cross-correlations}
\label{sec:21cm_LAE_cross_correlations}

The 21-cm line is emitted when a neutral hydrogen atom in its electronic ground state transitions from the triplet to the singlet hyperfine state. The spin temperature $T_s$ describes the ratio of atoms in the triplet to singlet state. It shapes the intensity of the emitted or absorbed 21cm radiation characterised by the brightness temperature $\delta T_b$. Importantly, we can only measure this 21cm radiation relative to the background radiation, the CMB, with a temperature $T_\mathrm{CMB}$. Thus the measurable differential 21cm brightness temperature $\delta T_b$ is given by \citep[e.g.][]{Furlanetto2006}
\begin{eqnarray}
\delta T_b(\bf{x}) &=& \frac{T_s({\bf x}) - T_\mathrm{CMB}}{1+z} \left( 1 - e^{-\tau({\bf x})}\right) \label{eq:deltaTb_exact} \\
&\simeq& \frac{3c\lambda_{21}^2 h A_{10}}{32 \pi k_\mathrm{B} H_0}\ n_{H,0} \left( \frac{1+z}{\Omega_m} \right)^{1/2} \left( 1-\frac{T_\mathrm{CMB}}{T_\mathrm{s}(\bf{x})} \right) \chi_\mathrm{HI}(\bf{x}) \nonumber \\
&=& T_0\  \left( 1-\frac{T_\mathrm{CMB}}{T_\mathrm{s}({\bf x})} \right) \chi_\mathrm{HI}({\bf x}) \left( 1 + \delta({\bf x}) \right)
\label{eq:deltaTb}
\end{eqnarray}
Here $\tau$ describes the corresponding 21cm optical depth, which we assume to be small in Eqn. \ref{eq:deltaTb}. $A_{10}$ represents the Einstein coefficient for spontaneous emission of a photon with an energy of $hc/\lambda_{21}$, corresponding to the energy difference of the hydrogen singlet and triplet hyperfine levels. $n_\mathrm{H,0}$ is the neutral hydrogen density today and $\chi_\mathrm{HI}({\bf x})$ and $1+\delta({\bf x})=\rho({\bf x})/\langle \rho\rangle$ describe the neutral hydrogen fraction and overdensity at position ${\bf x}$, respectively.

In this paper, we use the following definitions for the 21cm signal,
\begin{eqnarray}
\delta_{21}(\bf{x}) &=& \frac{\delta T_b(\bf{x})}{T_0} = \left( 1-\frac{T_\mathrm{CMB}}{T_s(\bf{x})}\right)\ \chi_\mathrm{HI}({\bf x}) \left( 1 + \delta(\bf{x}) \right), 
\label{eq:delta_21}
\end{eqnarray}
the number density of LAEs,
\begin{eqnarray}
\delta_\mathrm{LAE}(\bf{x}) &=& \frac{n_\mathrm{LAE}(\bf{x})}{\langle n_\mathrm{LAE}\rangle} - 1, 
\label{eq:delta_LAE} 
\end{eqnarray}
and the 21cm cross-correlation function, 
\begin{eqnarray}
\xi_\mathrm{21,LAE}(\bf{r}) &=& \frac{1}{V} \ \int \mathrm{d}^3x\ \ \delta_{21}(\bf{x}+\bf{r}) \ \delta_\mathrm{LAE}(\bf{x}), 
\end{eqnarray}
to derive the cross-correlations between the 21cm signal fluctuations and the LAE distribution.
In the following, we will phrase our calculations under the assumption that the simulation volume $V$ is gridded on $N$ cells, i.e. $\frac{1}{V}\int \mathrm{d}^3x \rightarrow \frac{1}{N} \sum_1^{N}$. While our calculations remain valid in the limit of $N\rightarrow\infty$, we choose this gridding approach to better reflect the typical outputs of simulations used to compute the 21cm-LAE cross-correlation functions.

\subsection{The cross-correlation amplitude at LAE positions}

To derive an analytic expression for the 21cm-LAE cross-correlation function, we first evaluate the 21cm-LAE cross-correlation function at the positions of LAEs,
\begin{eqnarray}
\xi_\mathrm{21,LAE}(r=0) &=& \frac{1}{N} \sum_{\bf{x}} \delta_{21}(\bf{x}) \ \delta_\mathrm{LAE}(\bf{x}), 
\end{eqnarray}
From Eqn. \ref{eq:delta_LAE} we see that $\delta_\mathrm{LAE}$ adopts only positive values at LAE locations and remains negative with a value of $-1$ otherwise, while the 21cm signal vanishes in ionised regions ($\chi_\mathrm{HI}=0$). 
LAEs are preferentially located in sufficiently large ionised regions with residual \HI fractions up to $10^{-4}$, allowing the Ly$\alpha$ line to redshift out of absorption and traverse the IGM. We note that sufficiently strong gas outflows from LAEs can relax this criterion, such that some LAEs could be located in neutral regions. However, as LAE surveys detect relatively bright LAEs \citep[e.g.][]{Ouchi2018} likely to be located in overdense and ionised regions, we assume here that $\delta_{21}=0$ at LAE locations for LAEs with $L_\alpha\gtrsim10^{42}$erg~s$^{-1}$. For these assumptions, the only regions contributing to $\xi_\mathrm{21,LAE}({\bf r}=0)$ are the neutral regions where no LAEs are found. Moreover, since the ionisation fronts are sharp, most of the $N$ cells will be either neutral or highly ionised ($\chi_\mathrm{HI}\lesssim10^{-4}$). We thus consider the ionisation field to be binary and neglect partially ionised cells at the ionisation fronts or around galaxies with ionised regions smaller than the cell size. The 21cm-LAE cross-correlation function values at very small scales are then given by
\begin{align}
\xi_\mathrm{21,LAE}(r=0) &= \frac{1}{N} \left[ \sum_{\bf{x}_\mathrm{HI}} \delta_{21}(\bf{x})\ \underbrace{\delta_\mathrm{LAE}(\bf{x})}_{=-1} +  \sum_{{\bf x}_\mathrm{HII}} \underbrace{\delta_{21}(\bf{x})}_{\simeq 0}\ \delta_\mathrm{LAE}(\bf{x}) \right] \nonumber \\
&= -\ \frac{1}{N} \sum_{{\bf x}_\mathrm{HI}}\ \left( 1-\frac{T_\mathrm{CMB}}{T_s(\bf{x})}\right)\ \underbrace{\chi_\mathrm{HI}({\bf x})}_{\simeq 1}\ \left(1+\delta(\bf{x})\right) \nonumber \\
&\simeq -\ \langle\chi_\mathrm{HI}\rangle\ \left< \left( 1-\frac{T_\mathrm{CMB}}{T_s} \right)\ \left( 1+\delta \right) \right>_\mathrm{HI}
\label{eq:xir0Ts}
\end{align}
Here $\langle\rangle_\mathrm{HI}$ denotes the mean value across neutral regions. We note that this expression for $\xi_\mathrm{21,LAE}(r=0)$ represents a lower limit, as LAEs in partially or complete neutral regions will contribute positively (based on the reasonable assumption that $T_s\gg T_\mathrm{CMB}$ at LAE locations).
As the Universe becomes ionised, the IGM is heated by the energetic photons from the first stars and galaxies, the spin temperature rises and exceeds the CMB temperature during the early phases of reionisation. Assuming the post-heating regime $T_\mathrm{s}\gg T_\mathrm{CMB}$ to be valid in neutral patches, the 21cm-LAE cross-correlation at very small scales becomes
\begin{eqnarray}
\xi_\mathrm{21,LAE}(r = 0) &\simeq& -\ \langle \chi_\mathrm{HI}\rangle\ \langle1+\delta\rangle_\mathrm{HI}
\label{eq:xir0}
\end{eqnarray}
We note that this limit also applies for any representation of $\delta_{21}$ that solely shifts the zero-point, e.g. $\delta_{21}({\bf x})=(\delta T_b({\bf x}) - \langle \delta T_b \rangle) / T_0$.

\subsection{The cross-correlation amplitude profile around LAEs}

Next we derive the 21cm-LAE cross-correlation profile depending on the size distribution of the ionised bubbles around LAEs. Here we limit our calculations to the post-heating regime of the EoR.
Separating the 21cm--LAE cross-correlation functions into $N_\mathrm{LAE}$ pixels containing ($\delta_\mathrm{LAE}>-1$) and $N-N_\mathrm{LAE}$ pixels devoid ($\delta_\mathrm{LAE}=-1$) of LAEs, we yield for the 21cm-LAE cross-correlation as a function of radial distance from an LAE
\begin{align}
\xi_\mathrm{21,LAE}(r) &= \frac{1}{N} \sum_{n=0}^{N} \delta_{21}({\bf x}+{\bf r})\ \delta_\mathrm{LAE}({\bf x}) \nonumber \\
&= \frac{1}{N} \sum_{n=0}^{N-N_\mathrm{LAE}} -\delta_{21}({\bf x}+{\bf r})|_{{\bf x}\neq{\bf x}_\mathrm{LAE}} \nonumber \\
& + \frac{1}{N} \sum_{n=0}^{N_\mathrm{LAE}} \frac{N}{N_\mathrm{LAE}} \delta_{21}({\bf x}+{\bf r})|_{{\bf x}={\bf x}_\mathrm{LAE}} \nonumber \\
&\simeq -\langle \delta_{21}({\bf x}) \rangle_{\bf x} + \frac{1}{N_\mathrm{LAE}}\sum_{n=0}^{N_\mathrm{LAE}} \delta_{21}({\bf x}+{\bf r})|_{{\bf x}={\bf x}_\mathrm{LAE}} \nonumber \\ 
&= -\langle \delta_{21}\rangle + \langle\delta_{21}\rangle^\mathrm{LAE}(r) \nonumber \\
&\simeq - \langle \chi_\mathrm{HI} \rangle \langle 1+\delta\rangle_\mathrm{HI} + \chi_\mathrm{HI}^\mathrm{LAE}(r) (1+\delta^\mathrm{LAE})(r).
\end{align}
Here we have assumed that pixels are either small enough to contain only one LAE or that LAEs are sparse enough that not more than one LAE is found in a pixel. $\langle\delta_{21}\rangle^\mathrm{LAE}$ is the average 21cm signal profile around LAEs, while $\langle \delta_{21}\rangle$ is the average overall 21cm signal. Correspondingly, $\chi_\mathrm{HI}^\mathrm{LAE}(r)$ and $(1+\delta^\mathrm{LAE})(r)$ are the average neutral fraction and density profiles around LAEs. 
\begin{eqnarray}
\xi_\mathrm{21,LAE}(r) &=& - \langle \chi_\mathrm{HI} \rangle \langle 1+\delta \rangle_\mathrm{HI} \left[ 1 - \frac{\chi_\mathrm{HI}^\mathrm{LAE}(r)}{\langle \chi_\mathrm{HI}\rangle} \frac{(1+\delta^\mathrm{LAE})(r)}{\langle 1 + \delta\rangle_\mathrm{HI}} \right] \nonumber \\
\label{eq:corrfunc21LAE_trend}
\end{eqnarray}
The main factor determining $\xi_\mathrm{21,LAE}(r)$ is the average neutral hydrogen profile around LAEs beyond the halo scale, $\chi_\mathrm{HI}^\mathrm{LAE}(r)$. While $\chi_\mathrm{HI}^\mathrm{LAE}(r)$ is determined by the sizes of the ionised regions around LAEs at small $r$ values, it converges to the average neutral hydrogen fraction $\langle\chi_\mathrm{HI}\rangle$ as $r$ increases beyond the typical sizes of the ionised regions around LAEs. 

To obtain an analytic form for $\xi_\mathrm{21,LAE}(r)$, we assume $(1+\delta^\mathrm{LAE})(r)\simeq \langle 1+\delta\rangle_\mathrm{HI}$ and derive $\chi_\mathrm{HI}^\mathrm{LAE}(r)$ as follows: We assume the distribution of the radii of the ionised regions around LAEs to follow a lognormal distribution \citep{Zahn2007, McQuinn2007, Meerburg2013}. With the probability density function
\begin{eqnarray}
\mathrm{PDF} (r) &=& \frac{1}{r\ \sqrt{2\pi \sigma_{\ln r}^2}} \exp\left[-\frac{\left[\ln \frac{r}{r_\mathrm{ion}}\right]^2}{2\sigma_\mathrm{ion}^2}\right],
\label{eq:PDF}
\end{eqnarray}
describing the probability of the size of an ionised region around an LAE, the cumulative density function 
\begin{eqnarray}
\mathrm{CDF} (r) &=& \int_0^{r} \mathrm{d}r'\ \mathrm{PDF}(r')\\
&=& \frac{1}{2} + \frac{1}{2}\mathrm{erf}\left[{\frac{\ln \frac{r}{r_\mathrm{ion}}}{\sqrt{2}\sigma_\mathrm{ion}}}\right]
\label{eq:CDF}
\end{eqnarray}
describes then the average profile of the neutral hydrogen fraction around LAEs, where each ionised region containing an LAE lies in an overall neutral medium, i.e. $\langle\chi_\mathrm{HI}\rangle\simeq1$. Inversely, $1-\mathrm{CDF}(r)$ depicts the average profile of the ionisation fraction around LAEs in an effectively neutral IGM. However, as the Universe becomes ionised, the distances between ionised regions reduces, and the probability to encounter a neutral or an ionised regions at large distances $r$ scales with the average neutral hydrogen fraction, $\langle\chi_\mathrm{HI}\rangle$. Therefore, we approximate the average neutral hydrogen fraction profile as
\begin{eqnarray}
\chi_\mathrm{HI}^\mathrm{LAE}(r) = \langle\chi_\mathrm{HI} \rangle ~ \mathrm{CDF}(r).
\end{eqnarray}
As the 21cm--LAE cross-correlation function depicts the probability of detecting a 21cm signal of a given strength at a distance $r$ from an LAE and thus traces the mean ionisation profile around LAEs, we propose the following ansatz for $\xi_\mathrm{21,LAE}$:
\begin{eqnarray}
\xi_\mathrm{21,LAE}(r) 
&=& -\ \langle\chi_\mathrm{HI}\rangle\ \langle1+\delta\rangle_\mathrm{HI}\ \left[ 1 - \mathrm{CDF}(r) \right] \nonumber \\
\label{eq:xi21LAE_analytic}
\end{eqnarray}
We will show in the following that this ansatz provides an excellent fit for the numerically derived results.

We note that $\chi_\mathrm{HI}^\mathrm{LAE}(r)$ may not be entirely dominated by the size distribution of the ionised regions around LAEs. At $r\lesssim 5$~cMpc self-shielding systems can increase the neutral hydrogen fraction around LAEs, leading to $\chi_\mathrm{HI}^\mathrm{LAE}(r\lesssim5~\mathrm{cMpc})>0$. At these distances, the average density profile around LAEs increases towards smaller distances. For example, in the {\sc astraeus} simulations, it rises approximately as $(1+\delta)(r)\simeq 1+ \frac{4}{3}r^{-4/3}$. The increase in the neutral hydrogen density towards LAEs causes then the 21cm-LAE cross-correlation function to reduce its negative amplitude towards smaller distances $r$ \citep[c.f. Eqn \ref{eq:corrfunc21LAE_trend} and][]{Weinberger2020, Kubota2018}. The presence of this feature depends strongly on the modelling of the self-shielding systems in simulations \citep[see e.g. Appendix D in][]{Weinberger2020}, which is a complex function of the ionising radiation and feedback processes from the stellar populations as well as the temperature and metallicity of the IGM gas.

\section{Results from simulations}
\label{sec:results_from_simulations}

In this Section we describe the different reionisation scenarios and simulations that we use then to analyse the dependency of the 21cm-LAE cross-correlation functions on reionisation and its topology. 

\subsection{Simulations}

We analyse results from two different reionisation simulation frameworks, (1) the semi-numerical galaxy evolution and reionisation model {\sc astraeus} and (2) the semi-numerical {\sc 21cmfast} code, which we describe in the following. Both use cosmological parameters consistent with the results from the Planck mission; the exact values used can be found in \citet{Klypin2016} for the underlying {\sc vsmdpl} simulation that {\sc astraeus} uses and \citet{Mesinger2016} for the EOS simulations.

\subsubsection{{\sc astraeus} simulations \citep{Hutter2022}}

{\sc astraeus}\footnote{\url{https://github.com/annehutter/astraeus}} couples a semi-analytical galaxy evolution model \citep[an enhanced version of {\sc delphi};][]{Dayal2014, Dayal2022, Mauerhofer2023} to a semi-numerical reionisation scheme \citep[{\sc cifog};][]{Hutter2018a} and runs on the outputs of a dark-matter (DM) only N-body simulation (merger trees and density fields). It includes not only models for all key processes of galaxy evolution thought to be relevant during the EoR, such as gas accretion, mergers, star formation, supernovae feedback, metal and dust enrichment, radiative feedback from reionisation, but also follows the spatially inhomogeneous reionisation process accounting for recombinations and tracking the residual \HI fraction in ionised regions \citep[see][for modelling details]{Hutter2021a, Ucci2022, Hutter2022}. 

The {\sc astraeus} simulations exploited for this analysis are based on the high-resolution {\sc very small multidark planck} ({\sc vsmdpl}) DM-only simulation from the {\sc multidark} simulation project and has been run with the {\sc gadget-2} Tree+PM N-body code \citep{Springel2005}. 
The {\sc vsmdpl} simulation follows the trajectories of $3840^3$ DM particles in a box with a side length of $160 h^{-1}$ comoving Mpc (cMpc), and each DM particle has a mass of $6.2 \times 10^6 h^{-1}\, \msun$. Halos and subhalos down to $20$ particles or a minimum halo mass of $1.24 \times 10^8h^{-1}\msun$ have been identified with the phase space {\sc rockstar} halo finder \citep{behroozi2013_rs} for all $150$ snapshots ranging from $z=25$ to $z=0$. 
To generate the necessary input files for {\sc astraeus}, we have used the pipeline internal {\sc cutnresort} scheme to cut and resort the vertical merger trees for $z=0$ galaxies (sorted on a tree-branch by tree-branch basis within a tree and generated by {\sc consistent trees}; \citet{behroozi2013_trees}) to local horizontal merger trees (sorted on a redshift-by-redshift basis within a tree) for galaxies at $z=4.5$. Moreover, for all snapshots at $z\geq4.5$, we have mapped the DM particles onto $2048^3$ grids and re-sampling these to $512^3$ grids to generate the DM density fields with cells with a side length of $312.5h^{-1}$~ckpc.

{\sc astraeus} has recently been extended to also include a model for Lyman-$\alpha$ emitters \citep[see][]{Hutter2022}, where the latter are defined as all galaxies exceeding a Ly$\alpha$ luminosity of $L_\alpha \geq 10^{42}$erg~s$^{-1}$. This model describes the Ly$\alpha$ line profile emerging from a galaxy as a function of the ISM gas and dust distribution as well as the escape fraction of \HI ionising photons ($f_\mathrm{esc}$), and follows the line-of-sight dependent Ly$\alpha$ attenuation by the \HI in the IGM during reionisation. In \citet{Hutter2022}, we explored three different Ly$\alpha$ line profile models and underlying relations between $f_\mathrm{esc}$ and halo mass. Here we will consider the two physically plausible bracketing reionisation scenarios of $f_\mathrm{esc}$ increasing ({\sc mhinc}) and decreasing ({\sc mhdec}) with rising halo mass, and the {\it Gaussian} Ly$\alpha$ line profile model also used in \citet{Hutter2014} and \citet{Dayal2011}. We note that these {\sc astraeus} simulations reproduce all available observational star-forming galaxy data sets at $z=5-10$, such as the ultraviolet luminosity functions, stellar mass functions, star formation rate and stellar mass densities. Moreover, the $f_\mathrm{esc}$ relations are normalised such that they reproduce the constraints on the reionisation history from GRB optical afterglow spectrum analyses, quasar sightlines, Ly$\alpha$ luminosity functions, Ly$\alpha$ emitter clustering and fraction as well as the CMB optical depth from \citet{Planck2020}.

For this combination of Ly$\alpha$ line profile model and reionisation scenarios, we compute the 21cm-LAE cross-correlation functions following the approach outlined in \citet{Hutter2017, Hutter2018b}. In brief, we derive the 21cm signal fields from the simulated ionisation ($\chi_\mathrm{HI}({\bf x})$) and density grids ($1+\delta({\bf x})$) by applying Eqn. \ref{eq:deltaTb}, assuming $T_x({\bf x}) >> T_\mathrm{CMB}$ and
\begin{eqnarray}
T_0 &=& 28.5\mathrm{mK} \left( \frac{1+z}{10} \right)^{1/2} \frac{\Omega_b}{0.042} \frac{h}{0.073} \left( \frac{\Omega_m}{0.24} \right)^{-1/2},
\end{eqnarray}
to each grid cell. We then obtain the dimensionless 21cm-LAE cross-correlation function as
\begin{eqnarray}
\xi_\mathrm{21,LAE}(r) &=& \int P_\mathrm{21,LAE}(k)\ \frac{\sin(kr)}{kr}\ 4\pi k^2\ \mathrm{d}k
\end{eqnarray}
The cross-power spectrum $P_\mathrm{21,LAE}(k) = V \langle \widetilde{\delta}_{21}({\bf k})\ \widetilde{\delta}_\mathrm{LAE}({\bf -k}) \rangle$ is in units of cMpc$^3$ for a volume $V$ and derived from the product of the Fourier transformation of the fractional fluctuation fields $\delta_{21}({\bf x})$ and $\delta_\mathrm{LAE}({\bf x})$ as defined in Eqn. \ref{eq:delta_21} and \ref{eq:delta_LAE}.\footnote{The Fourier transformation is computed as $\widetilde{\delta}({\bf k}) = V^{-1} \int \delta({\bf x})\ e^{-2\pi i {\bf kx}}\ \mathrm{d}^3x$.}
We note that the 21cm-LAE cross-correlation results for the {\it Clumpy} and {\it Porous} Ly$\alpha$ line profile models also explored in \citet{Hutter2022} are identical to those of the {\it Gaussian} model, as the galaxies identified as observable LAEs, i.e. after accounting for the attenuation by the IGM, are effectively the same. While the {\it Gaussian} Ly$\alpha$ line profile model describes the Ly$\alpha$ line emerging from galaxies as a Gaussian centred around the Ly$\alpha$ resonance, the {\it Clumpy} and {\it Porous} models consider the gas and dust to be clumpy and in case of the {\it Porous} model also dispersed with gas-free channels, resulting in double-peak profiles with varying emission at the Ly$\alpha$ resonance depending on the assumed clump size and $f_\mathrm{esc}$.

\subsubsection{EOS {\sc 21cmfast} simulations \citep{Mesinger2016}}

{\sc 21cmfast} combines the excursion-set formalism and perturbation theory to follow the evolving density, velocity, ionisation, and spin temperature fields. The Evolution of 21cm Structure (EOS) project\footnote{\url{http://homepage.sns.it/mesinger/EOS.html}} provides public 21cm simulations of the EoR of 1.6 Gpc box length, computed on a 1024$^3$ grid using {\sc 21cmfastv2} \citep{sobacchi14}, and cell sizes of $\sim1 h^{-1}$~Mpc. The EOS simulations include a sub-grid prescription for inhomogeneous recombinations, photo-heating suppression of the gas fraction in small halos, and a calibration of the X-ray emissivity of galaxies with high-mass X-ray binary observations in local star forming galaxies \citep{Mineo12}. Also, the Lyman series radiation background is self-consistently computed, determining how closely the spin temperature tracks the kinetic gas temperature through the Wouthuysen-Field effect \citep{Wouthuysen1952, Field1958}. The EOS simulations explore two models for the EoR morphology: (1) the faint galaxy model characterised by many small ionised HII regions (SmallHII), and (2) the bright galaxy model of fewer, larger HII regions (LargeHII). These two models are based on different star-formation scenarios, corresponding to efficient star formation in either faint or bright galaxies and, thus, different typical masses for the underlying dark matter halos. In both cases, the (constant) ionising escape fraction is matched to yield similar Thompson scattering optical depths, consistent with estimates from {\it Planck} \citep{Planck2016}. To assign LAEs to host halos, we connect LAE intrinsic luminosity and host halo mass via a minimum halo mass (corresponding to a minimum observed luminosity) as well as a duty cycle that accounts for the stochasticity of Ly$\alpha$ emission. This relation is calibrated to match the observed $z=6.6$ LAE number density and luminosity function of the Subaru Suprime-Cam ultra-deep (UD) field \citep{Ouchi2010}, taking into account the IGM attenuation along the line-of-sight for a typical velocity shift of $\sim 230$km/s redward of the line center. We define LAEs as galaxies with a Ly$\alpha$ luminosity of $L_\alpha\geq2.5\times10^{42}$erg~s$^{-1}$ and have checked that the resulting LAEs also match the observed angular clustering signal.

We calculate the cross-correlation function directly from our real-space 21cm and LAE boxes using the estimator from \citet{Croft16}. We do not directly Fourier transform from the cross-power spectrum to the cross-correlation function, as we found this to be less stable in the presence of 21cm noise in mock realisations. Our noise model assumes an SKA1-low tracked scanning strategy with 1000h on-sky integration and is calculated using the 21cmSense code \citep{Pober13,Pober14}. Specifically, we assume modes in the so-called foreground wedge to be lost, a frequency-dependent scaling for the sky temperature, and a compact antennae core of a maximum baseline of 1.7 km for the antennae configuration from the SKA1-low baseline design. We sum over the visible, narrow-band projected LAE--21cm cell pairs at distance $r$, 
\begin{equation}
r_\mathrm{21,LAE} \left( r \right) = \frac{1}{N_\mathrm{LAE}N(r)} \sum_i^{N_\mathrm{LAE}} \sum_j^{N(r)} \delta_{21}\left(\textbf{r}_{\rm i}+\textbf{r}_{\rm j}\right)~,
\label{eq:r21LAE}
\end{equation}
where $\textbf{r}_{\rm i}$ is the position of the $i$-th LAE and $|\textbf{r}_{\rm j}|=r$; $N_{\rm LAE}$ is the number of LAEs in the observed volume and $N\left(r\right)$ is the number of 21cm cells at distance $r$ from the i-th LAE.

\begin{figure*}
    \centering
         \includegraphics[width=\textwidth]
         {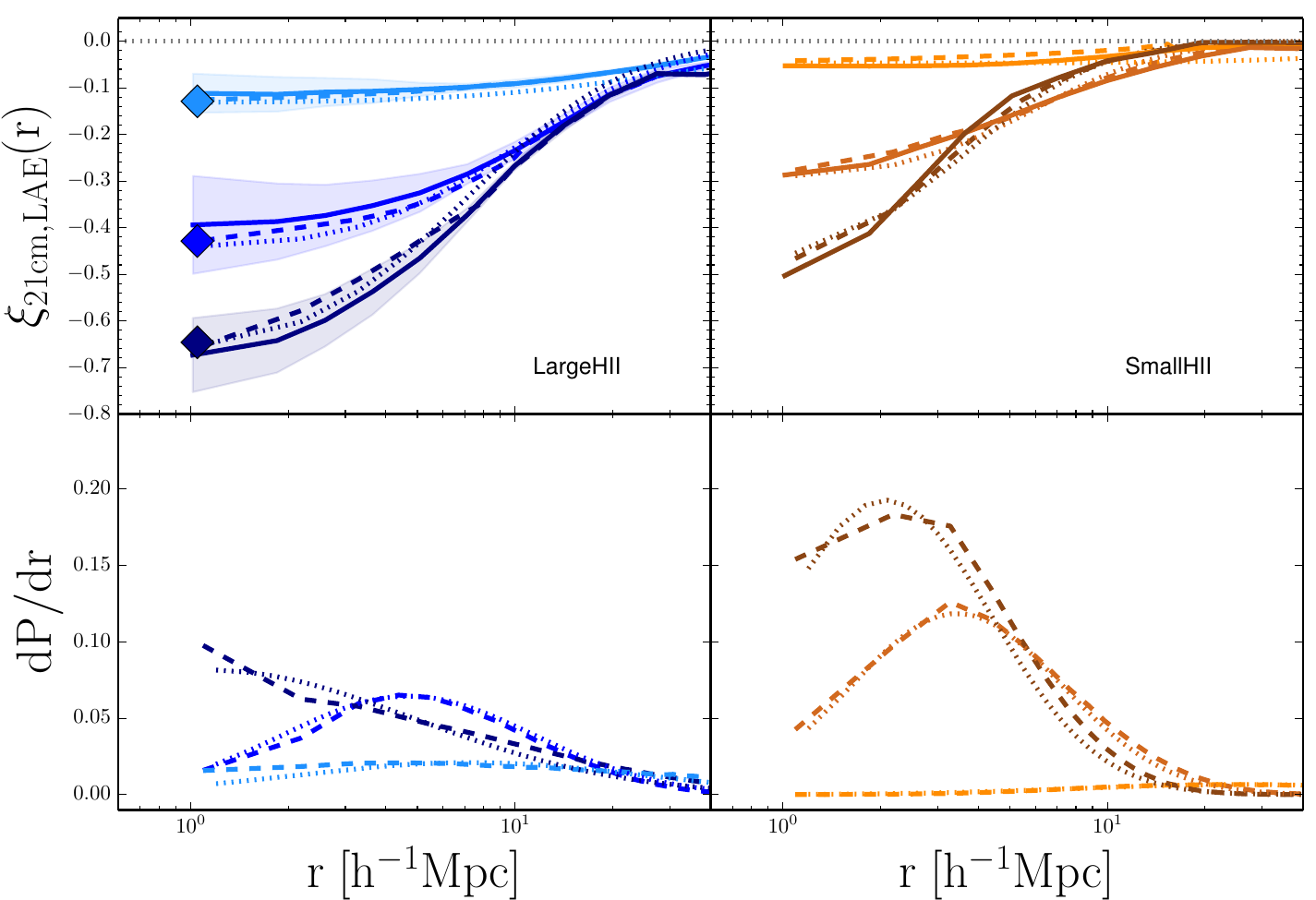}
    \caption{
    Upper panels: 21cm -- LAE cross-correlation functions for the EOS LargeHII (left) and SmallHII (right) simulations for varied hydrogen neutral fraction, for neutral hydrogen fraction $\sim 0.15$, $\sim0.52$ and $\sim 0.74$ (top to bottom line, light to dark). Shaded regions depict 2$\sigma$ scatter computed for each 10 mock Monte-Carlo SKA1-low and Subaru HSC realisations. The diamonds at small $r$ (left) depict the $r=0$ Eqn.\eqref{eq:xir0} expectation as derived from the corresponding simulations. Bottom: Probability density distribution of ionised regions of the LargeHII (left) and SmallHII (right) simulations. Solid lines show the results from the simulations in the top panels. Dashed and dotted lines show our analytical fit using the (1) ionisation profiles, and (2) the lognormal distribution, respectively. }
    \label{fig:crossr0}
\end{figure*}

\begin{figure}
    \centering
    \includegraphics[width=\columnwidth]{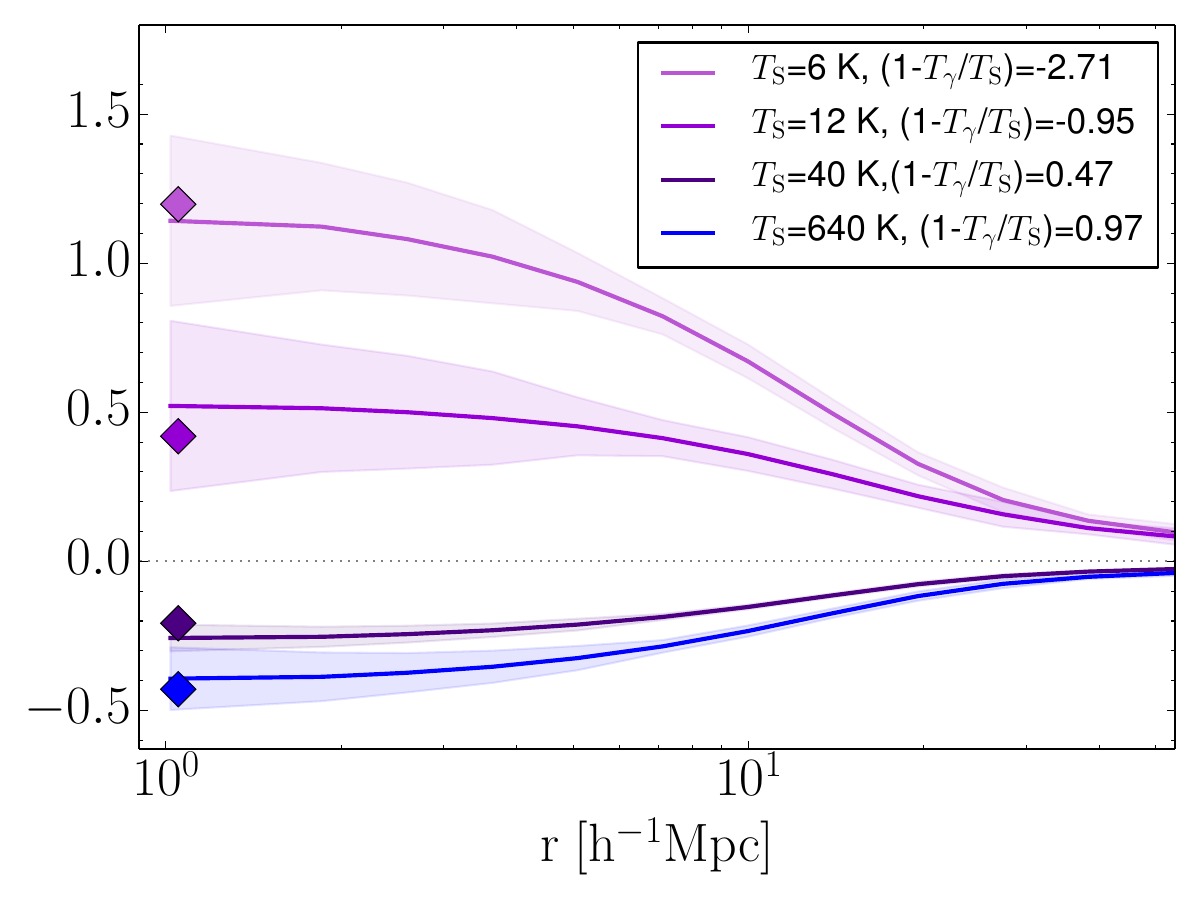}
    \caption{
    21cm -- LAE cross-correlation functions for the EOS LargeHII and different average spin temperature values $T_\mathrm{s}$ in the neutral IGM at fixed neutral fraction of $\sim 50 \%$. Shaded regions depict 2$\sigma$ scatter computed for each 10 mock Monte-Carlo SKA1-Low (1000h) and Subaru HSC realisations. }
    \label{fig:crossrTs}
\end{figure}

\subsection{Understanding the 21cm-LAE cross-correlation dependencies}

\begin{table}
    \centering
    \begin{tabular}{c|c|c|c|c|c|c}
        \hline
        \hline
         & {\sc mhinc} & {\sc mhinc} & {\sc mhinc} & {\sc mhdec} & {\sc mhdec} & {\sc mhdec} \\
        \hline
        $z$ & $\langle\chi_\mathrm{HI}\rangle$ & $\frac{r_\mathrm{ion}}{h^{-1}\mathrm{cMpc}}$ & $\frac{\sigma_\mathrm{ion}}{h^{-1}\mathrm{cMpc}}$ & $\langle\chi_\mathrm{HI}\rangle$ & $\frac{r_\mathrm{ion}}{h^{-1}\mathrm{cMpc}}$ & $\frac{\sigma_\mathrm{ion}}{h^{-1}\mathrm{cMpc}}$ \\
        \hline
        \hline
        8.0 & 0.84 & 3.70 & 0.70 & 0.71 & 3.45 & 0.66 \\
        7.3 & 0.69 & 5.68 & 0.81 & 0.59 & 4.68 & 0.72 \\
        7.0 & 0.52 & 7.24 & 0.96 & 0.49 & 5.48 & 0.80 \\
        6.6 & 0.23 & 13.00 & 1.14 & 0.34 & 8.19 & 0.97 \\
        \hline
    \end{tabular}
    \caption{Best fit values for fitting the lognormal distribution to the size distribution of ionised regions around LAEs derived from the {\sc astraeus} {\sc mhinc} and {\sc mhdec} simulations. In the {\sc mhinc} simulation, $f_\mathrm{esc}$ increases with rising halo mass, while it decreases in the {\sc mhdec} simulation.}
    \label{tab:astraeus}
\end{table}

\begin{table}
    \centering
    \begin{tabular}{c|c|c|c|c}
        \hline
        \hline
         & LargeHII & LargeHII & SmallHII & SmallHII \\
        \hline
        $\langle\chi_\mathrm{HI}\rangle$ & $\frac{r_\mathrm{ion}}{h^{-1}\mathrm{cMpc}}$ & $\frac{\sigma_\mathrm{ion}}{h^{-1}\mathrm{cMpc}}$ & $\frac{r_\mathrm{ion}}{h^{-1}\mathrm{cMpc}}$ & $\frac{\sigma_\mathrm{ion}}{h^{-1}\mathrm{cMpc}}$ \\
        \hline
        \hline
        0.74 & <9.77$^{*}$ & <1.47$^{*}$ & 3.65 & 0.75 \\
        0.52 & 10.33 & 0.89 & 5.97 & 0.74 \\
        0.16 / 0.15\tablefootnote{During late states of reionisation, or at comparably low neutral hydrogen fraction, the size distribution of ionised regions in the EOS simulations has a broad peak at a few tens of Mpc with a considerable tail towards larger radii, traceable due to the large simulated volume of 1$\,$Gpc$^3$. We therefore caution the best fit values in this row to have a large uncertainty.} & 32.81 & 1.23& 100.61 & 1.16 \\
        \hline
    \end{tabular}
    \caption{Best fit values for fitting the lognormal distribution to the size distribution of ionised regions around LAEs derived from the LargeHII and SmallHII model of the EOS simulations.
    \newline $^{*}$ Upper limit due to limited spatial resolution of the simulations.}
    \label{tab:EOS}
\end{table}

\begin{table}
    \centering
    \begin{tabular}{c|c|c|c}
        \hline
        \hline
        $\langle\chi_\mathrm{HI}\rangle$ & $\langle 1 + \delta \rangle_\mathrm{HI}$ for LargeHII & $\langle 1 + \delta \rangle_\mathrm{HI}$ for SmallHII \\
        \hline
        \hline
        0.74 & 0.90 & 0.89 \\
        0.52 & 0.85 & 0.83 \\
        0.16 / 0.15 & 0.83 & 0.78 \\
        \hline
    \end{tabular}
    \caption{Neutral gas overdensities at given global neutral hydrogen fractions for the LargeHII and SmallHII simulations.}
    \label{tab:EOS_XHI_neutral_overdensity}
\end{table}

\begin{figure*}
    \centering
    \includegraphics[width=\textwidth]{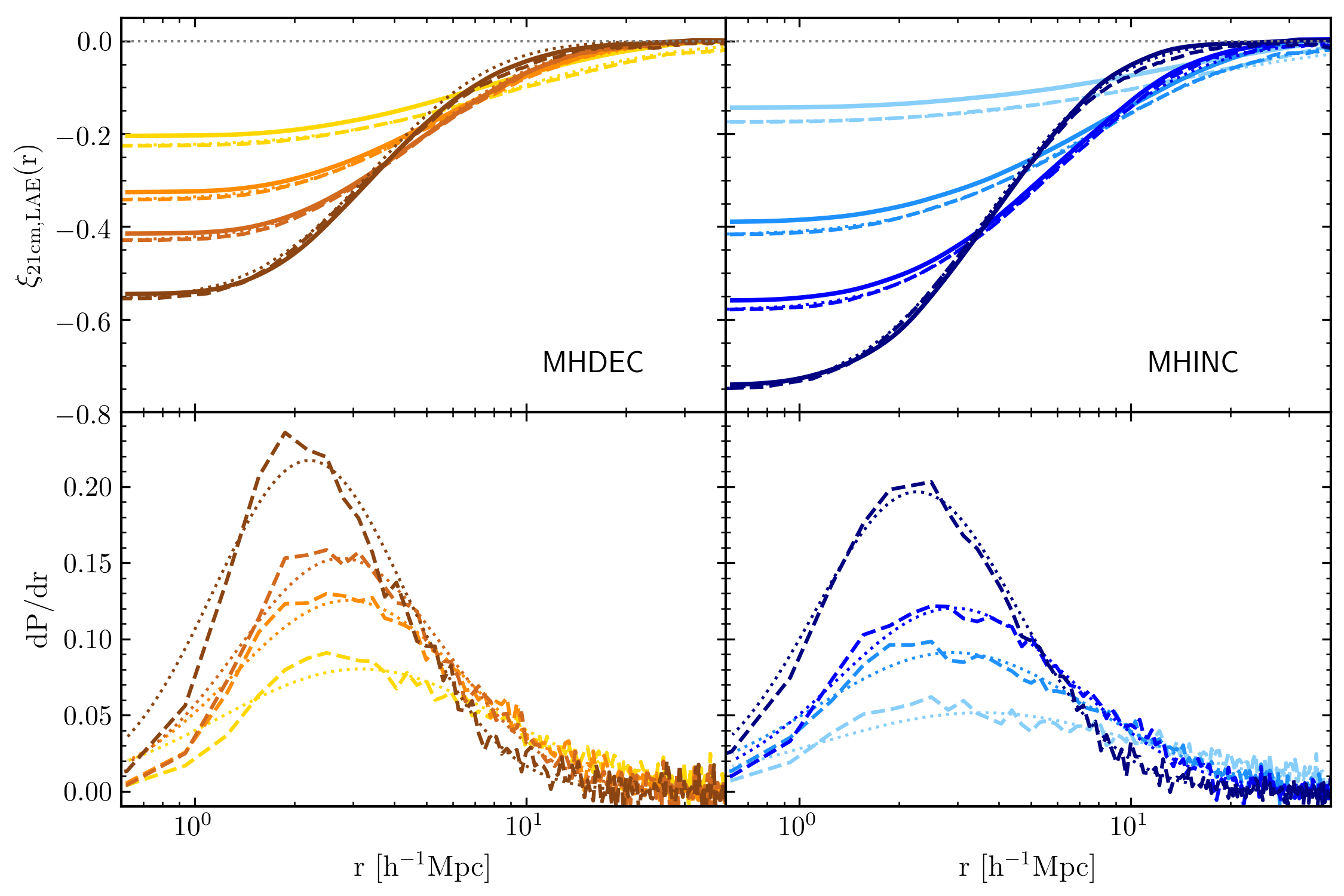}
    \caption{21cm-LAE cross-correlation functions (top) and probability density distribution of ionised regions (bottom) of the {\sc astraeus} {\sc mhdec} and {\sc mhinc} simulations  for varied global \HI fractions at $z=8.0$, $7.3$, $7.0$, $6.7$ from dark to bright colours. For these redshifts the \HI fractions are $\langle \chi_\mathrm{HI} \rangle=0.84$, $0.69$, $0.52$, $0.23$ for {\sc mhinc} and $0.71$, $0.59$, $0.49$, $0.34$ for the {\sc mhdec} simulations, respectively (see Tab. \ref{tab:astraeus}). Solid lines show the results from the simulations in the top panels. Dashed and dotted lines show our analytical fit using the (1) the ionisation profiles along the $6$ lines of sights (along major axes), and (2) the lognormal distribution that fits best to the line-of-sight averaged ionisation profile, respectively.}
    \label{fig:21cmLAEcrosscorrfunc_astraeus}
\end{figure*}

To understand how the 21cm-LAE cross-correlation function, particularly its amplitude, depends on the ionisation state of the IGM and the reionisation topology, we compare the 21cm-LAE cross-correlation functions derived from our simulations to the analytic limits and profiles outlined in Section \ref{sec:21cm_LAE_cross_correlations}. We show the respective 21cm-LAE cross-correlation functions at different stages of reionisation and for different scenarios in Fig. \ref{fig:crossr0} for the EOS simulations (SmallHII, LargeHII) and in Fig. \ref{fig:21cmLAEcrosscorrfunc_astraeus} for the {\sc astraeus} simulations ({\sc mhdec}, {\sc mhinc}). 

Firstly, from these figures, we see that the negative 21cm-LAE cross-correlation amplitude at small scales, $|\xi_\mathrm{21,LAE}(r=0)|$, decreases in all reionisation scenarios as the Universe becomes more ionised. This trend has been found in a number of works \citep[e.g.][]{Sobacchi2016, Hutter2017, Heneka2017, Hutter2018b, Heneka2020, Weinberger2020} and agrees with the $\langle\chi_\mathrm{HI}\rangle$-scaling of $\xi_\mathrm{21,LAE}$ derived in Eqn. \ref{eq:xir0} and \ref{eq:corrfunc21LAE_trend}. The latter echos the fact that LAEs are located in ionised regions, and thus the difference between the average ionisation level and that at LAE positions is $\langle1-\chi_\mathrm{HII}\rangle = \langle\chi_\mathrm{HI}\rangle$. 

However, as we expect from Eqn. \ref{eq:xir0} and \ref{eq:corrfunc21LAE_trend} and can see from the simulated $\xi_\mathrm{21,LAE}$ values and the corresponding ionisation levels (c.f. Tables \ref{tab:astraeus} and \ref{tab:EOS}), the overall ionisation state of the IGM is not the only quantity that defines $|\xi_\mathrm{21,LAE}(r=0)|$. As the average 21cm differential brightness temperature depends on the ionisation state {\it and} the gas density in neutral regions, $|\xi_\mathrm{21,LAE}(r=0)|$ is also proportional to the gas overdensity in neutral regions, $\langle 1+\delta\rangle_\mathrm{HI}$. 
In Fig.~\ref{fig:crossr0} and \ref{fig:21cmLAEcrosscorrfunc_astraeus}, we see that the $|\xi_\mathrm{21,LAE}(r=0)|$ expectations according to Eqn. \ref{eq:xir0} (depicted as diamonds in Fig. \ref{fig:crossr0} and dotted lines in Fig. \ref{fig:21cmLAEcrosscorrfunc_astraeus}) match well with the simulation-derived cross-correlations at small $r$ (solid lines) for the LargeHII, {\sc mhinc} and {\sc mhdec} models. 
We note that for the SmallHII model the mean neutral density and thus the $|\xi_\mathrm{21,LAE}(r=0)|$ expectation is up to $\sim 10 \%$ lower in absolute value depending on $\langle\chi_\mathrm{HI}\rangle$ as compared to the LargeHII expectation; due to the on average smaller size of ionised regions in the SmallHII model we probably need to resolve smaller scales, such as in the {\sc astraeus} simulations ($\lesssim0.5h^{-1}$cMpc), for a better extrapolation to $r=0$. The limits given are thus representing results for an upper limit on the bubble sizes. 

The shaded regions in Fig.~\ref{fig:crossr0} (as in Fig.~\ref{fig:crossrTs}) show the $2\sigma$ uncertainty from 10 Monte-Carlo mock realisations of the 21cm signal assuming 1000h of SKA-Low observations and of a narrow-band LAE survey with Subaru HSC characteristics. For the narrow-band LAE survey we assumed a systemic redshift uncertainty of $\Delta z = 0.1$, a survey area of $3.5$deg$^2$, and a limiting narrow-band luminosity of $L_{\alpha,\mathrm{min}} = 2.5 \times 10^{42}\,$erg$\,$s$^{-1}$. As can be seen in this figure, we can expect the cross-correlation signals at different neutral hydrogen fractions (0.15, 0.52, 0.74) depicted to be distinguishable with such experiments. We would like to draw attention here mostly to the finding, that our analytical expectation and the simulation-derived cross-correlations agree well within the uncertainty bands depicted.  

Secondly, we note that Eqn. \ref{eq:xir0} is only valid in the post-heating regime where $T_\mathrm{s}\gg T_\mathrm{CMB}$. During Cosmic Dawn when $T_\mathrm{s}\lesssim T_\mathrm{CMB}$, $\xi_\mathrm{21,LAE}(r=0)$ depends also on the average spin temperature in neutral regions (as $1-T_\mathrm{s}/T_\mathrm{CMB}$) as predicted by Eqn. \ref{eq:xir0Ts}. Fig. \ref{fig:crossrTs} depicts the 21cm-LAE cross-correlation function at $\langle\chi_\mathrm{HI}\rangle\simeq0.5$ for different average spin temperature values $T_\mathrm{s}$ in the neutral IGM derived from the EOS simulations (solid lines). These simulations track the spatially inhomogeneous evolution of the IGM temperature and Lyman series radiation background relevant for determining the coupling between the kinetic gas and spin temperatures, and follow the spin temperature fluctuations. In Fig. \ref{fig:crossrTs}, the spatial fluctuations of the spin temperature were considered when calculating the 21cm brightness temperature and respective cross-correlations. We refer the reader to \citet{Heneka2020} for a detailed discussion of how the spin temperature fluctuations shape the 21cm-LAE cross-correlation functions when the IGM or parts of it remain cold, revealing that assuming an average spin temperature would not yield the same results. The comparison to the analytical prediction at $r\simeq0$ (coloured diamonds) shows again that these are in good agreement with the results from the simulations. This underlines that $\xi_\mathrm{21,LAE}$ is sensitive to the spin temperature and gas densities in neutral regions and not to their full-box averages. We explore this power of the 21cm signal to probe the density in neutral regions in the next Section.

\begin{figure}
    \centering
    \includegraphics[width=0.5\textwidth]{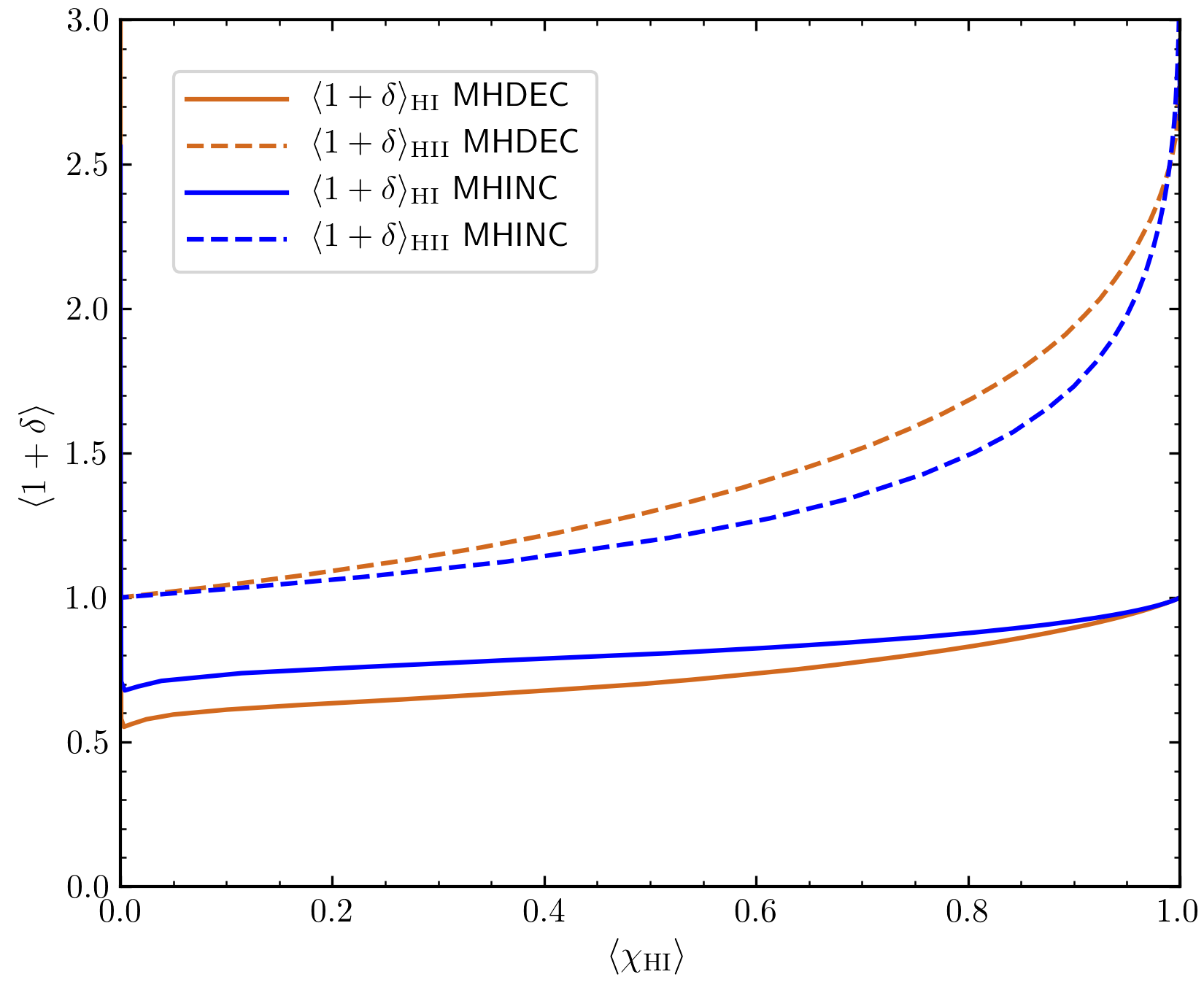}
    \caption{Relation between the neutral (ionised) gas overdensity and the global neutral hydrogen fraction in the {\sc astraeus} {\sc mhdec} and {\sc mhinc} simulations. Solid lines show the results for the neutral hydrogen gas and dashed lines for the ionised hydrogen gas. }
    \label{fig:dens_XHI_relation_astraeus}
\end{figure}

Next, we analyse how the size distribution of the ionised regions around LAEs is imprinted in $\xi_\mathrm{21,LAE}(r)$. For this purpose, we derive the size distribution of ionised regions around LAEs in both {\sc astraeus} and EOS simulations by shooting rays from each simulated LAEs along all major axes of the simulation box and measure the sizes of the surrounding ionised regions. Taking the resulting size distribution of ionised regions as $\mathrm{PDF}(r)$ (dashed lines in bottom panels of Fig. \ref{fig:crossr0} and Fig. \ref{fig:21cmLAEcrosscorrfunc_astraeus}), we derive $\xi_\mathrm{21,LAE}(r)$ with Eqn. \ref{eq:CDF} and \ref{eq:xi21LAE_analytic} (dashed lines in top panels of Fig. \ref{fig:crossr0} and Fig. \ref{fig:21cmLAEcrosscorrfunc_astraeus}). The derived $\xi_\mathrm{21,LAE}(r)$ values agree very well with the numerically derived ones. Various works have pointed out that the 21cm-LAE cross-correlation function can measure the typical sizes of ionised regions around LAEs \citep[e.g.][]{Lidz2009, Wiersma2013, Vrbanec2020}, however they do not agree on which characteristic point in $\xi_\mathrm{21,LAE}(r)$ traces the scale of the average or typical size of ionised regions. Here we confirm that the peak of the size distribution of the ionised regions coincides with the inflection point of $\xi_\mathrm{21,LAE}(r)$. We also note that at the same $\langle \chi_\mathrm{HI} \rangle$ values the ionised regions around LAEs in the EOS simulations have on average larger sizes than in the {\sc astraeus} simulations (c.f. Tables \ref{tab:astraeus} and \ref{tab:EOS} at $\langle \chi_\mathrm{HI} \rangle \simeq0.5$). This might be due to the {\sc astraeus} simulations assuming a lower Ly$\alpha$ luminosity required for a galaxy to be an LAE and an ionising emissivity biased more towards lower-mass halos through the scaling of $f_\mathrm{esc}$ with halo mass.

Finally, we test whether a lognormal distribution adequately describes the size distribution of the ionised regions around LAEs and can be used to quickly forecast $\xi_\mathrm{21,LAE}(r)$ from a given set of parameters ($\langle\chi_\mathrm{HI}\rangle$, $\langle 1+\delta\rangle_\mathrm{HI}$, $r_\mathrm{ion}$, $\sigma_\mathrm{ion}$). We test this hypothesis by finding the parameters of the lognormal distribution ($r_\mathrm{ion}$, $\sigma_\mathrm{ion}$) that best fit the $\mathrm{PDF}(r)$ derived from the measured ionisation profiles around LAEs in the simulations. The bottom panels of Fig. \ref{fig:21cmLAEcrosscorrfunc_astraeus} show that the lognormal distribution (dotted lines) provides indeed a good fit to the measured size distributions of ionised regions (dashed lines). Most notably the lognormal distribution only tends to overpredict the number of smaller sized ionised regions, leading to the corresponding $\xi_\mathrm{21,LAE}(r)$ values (derived following Eqn. \ref{eq:xi21LAE_analytic} and shown as dashed lines in top panels of Fig. \ref{fig:21cmLAEcrosscorrfunc_astraeus}) shifting to higher values or smaller scales.

In summary, we find our analytic limits and profiles to match the $\xi_\mathrm{21,LAE}(r)$ values derived from the EOS and {\sc astraeus} simulations very well as long as LAEs reside in highly ionised cells. Assuming a lognormal distribution is an adequate approximation for the size distribution of ionised regions and can be used to fit future 21cm-LAE cross-correlation functions derived from observations.

\subsection{Tracing the reionisation topology}

The different reionisation scenarios covered in the EOS and {\sc astraeus} simulations allow us to analyse the signatures of the reionisation topology in their 21cm-LAE cross-correlation function. We focus on two signatures: the small-scale amplitude and the inflection point of the 21cm-LAE cross-correlation function $\xi_\mathrm{21,LAE}(r)$.

Firstly, the small-scale amplitude $|\xi_\mathrm{21,LAE}|(r\simeq0)$ depends on the reionisation topology, i.e. the propagation of the ionisation fronts through the cosmic web, as it traces the average hydrogen gas density in neutral regions, $\langle1+\delta\rangle_\mathrm{HI}$: the stronger the correlation between the underlying gas distribution and ionising emissivity distribution emerging from galaxies or the redshift when a region became ionised, the lower is $\langle 1+\delta\rangle_\mathrm{HI}$ at any time during the EoR. We see this relation when comparing the SmallHII and LargeHII reionisation scenarios in the EOS simulations in Tab. \ref{tab:EOS_XHI_neutral_overdensity} and the {\sc mhdec} and {\sc mhinc} reionisation scenarios in the {\sc astraeus} simulations in Fig. \ref{fig:dens_XHI_relation_astraeus}. 
In both, the LargeHII and {\sc mhinc} scenario, the majority of the ionising photons are produced and escape from more massive galaxies with $T_\mathrm{vir}>2\times10^5$K and $M_h\gtrsim10^{9.5}\msun$, respectively. Located in significantly overdense regions, the ionised regions originating from these galaxies trace indeed these significantly overdense regions but not the less dense regions where lower mass halos are located; $\langle1+\delta\rangle_\mathrm{HI}$ drops only slightly as reionisation progresses. In contrast, ionised regions in the SmallHII and {\sc mhdec} scenarios follow the underlying DM and gas density distribution closely as the low-mass halos located in intermediate to dense regions are the dominant sources of ionising photons; as a consequence $\langle1+\delta\rangle_\mathrm{HI}$ traces increasingly the least dense regions as the Universe becomes ionised.

Secondly, the shape of $\xi_\mathrm{21,LAE}$ directly maps the size distribution of the ionised regions; in particular, the peak of the size distribution of ionised regions coincides with the inflection point of $\xi_\mathrm{21,LAE}$.\footnote{The inflection point is given by $\frac{\partial^2 \xi_\mathrm{21,LAE}(r)}{\partial r^2} \simeq \frac{\partial^2 \mathrm{CDF}(r)}{\partial r^2} = \frac{\partial \mathrm{PDF}(r)}{\partial r} = 0$.} Importantly, the peak of the size distribution is highly sensitive to the distribution of the ionising emissivity within the galaxy population, e.g. the more ionising radiation escapes from lower mass halos, the more similar sized are the ionised regions and the smaller is the average ionised region. Indeed, these trends can be seen in Fig. \ref{fig:21cmLAEcrosscorrfunc_astraeus} and Tab. \ref{tab:astraeus} when going from the {\sc mhinc} (bottom right panel) to the {\sc mhdec} scenario (bottom left panel): the size distribution of ionised regions becomes more peaked and shifts to smaller scales. Hence, the inflection point of $\xi_\mathrm{21,LAE}$ provides an estimate of the typical size of ionised regions around LAEs.

We note that the {\sc astraeus} simulations show lower $\langle1+\delta\rangle_\mathrm{HI}$ values than the EOS simulations due to the following reasons: (1) while both the SmallHII and {\sc mhdec} scenarios consider halos that exceed virial temperatures of $T_\mathrm{vir}=10^4$K and are not star-formation suppressed by radiative feedback from reionisation, the contribution of low-mass halos ($M_h\lesssim10^{9.5}\msun$) to reionisation is higher in the {\sc mhdec} simulation, as it includes also an ionising escape fraction ($f_\mathrm{esc}$) that decreases with rising halo mass. (2) While the LargeHII simulation considers only halos with $T_\mathrm{vir}>2\times10^5$K ($M_h\gtrsim10^{9.5}\msun$ at $z=7$) to contribute to the ionising budget, the {\sc mhinc} scenario includes the same halos as the {\sc mhdec} scenario but an $f_\mathrm{esc}$ that increases with halo mass and thus has also minor contribution from low-mass halos. 

In summary, as the reionisation topology depends sensitively on the trends of galactic properties shaping the emerging ionising emissivity with galaxy mass (e.g. $f_\mathrm{esc}$, stellar populations, initial mass function), not only the inflection point of the 21cm-LAE cross-correlation function traces the ionising properties of LAEs but also their small-scale amplitude during the EoR. The more ionising radiation emerges from low-mass objects that follow the underlying cosmic web structure more closely than more massive objects, the stronger is the correlation between the underlying density and ionisation fields,\footnote{We note that a stronger correlation between the underlying density and ionisation fields results in a lower average overdensity in neutral regions $\langle 1+\delta\rangle_\mathrm{HI}$ at fixed $\langle\chi_\mathrm{HI}\rangle$.} and thus the weaker is the 21cm-LAE anti-correlation amplitude at small scales.

\begin{figure*}
    \centering
    \includegraphics[width=0.99\textwidth]{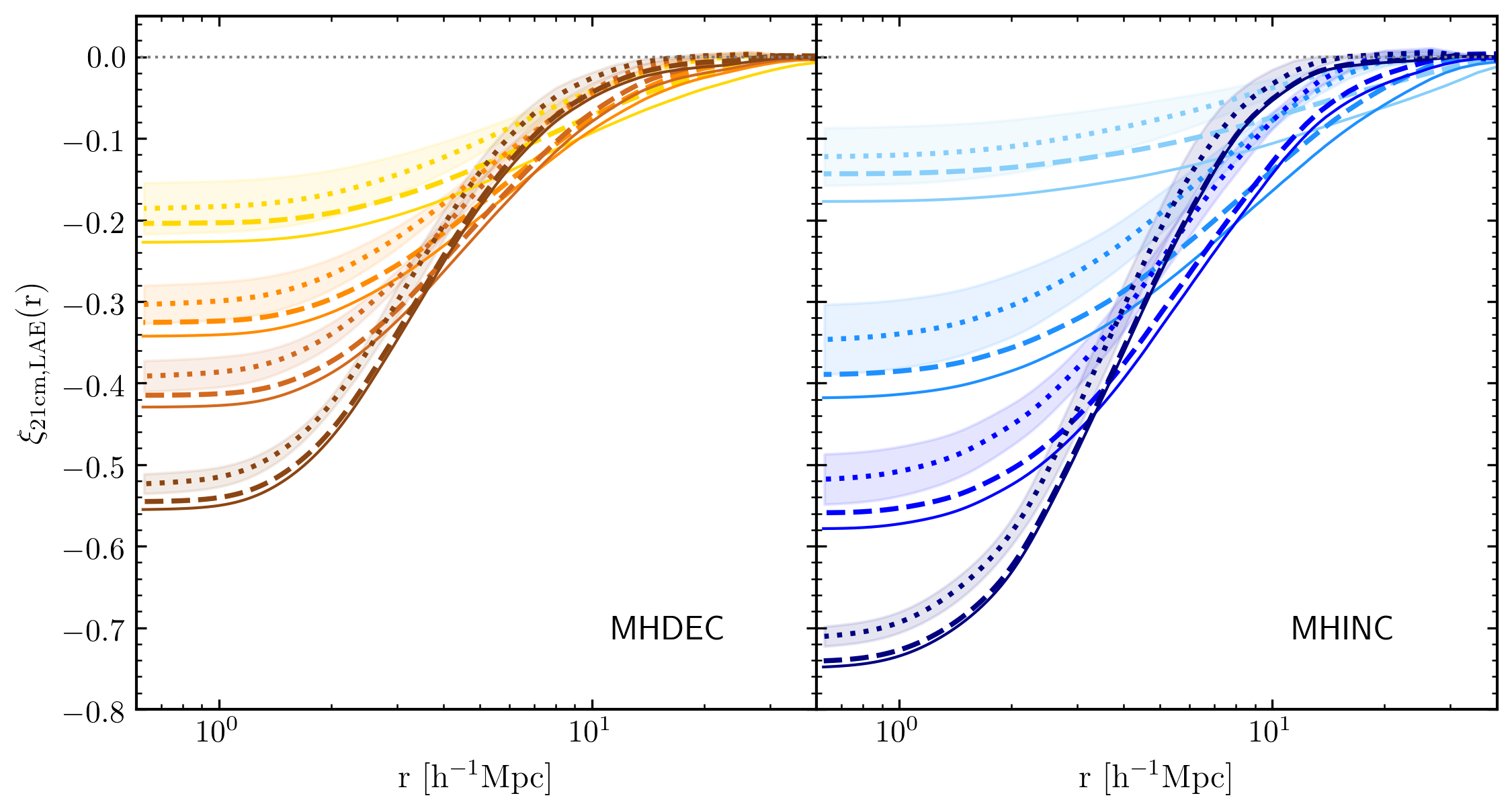}
    \caption{21cm-LAE cross-correlation functions of the {\sc astraeus} {\sc mhdec} and {\sc mhinc} simulations for simulation box sizes of $80h^{-1}$cMpc (dotted), $160h^{-1}$cMpc (dashed), and $320h^{-1}$cMpc (solid). The shaded region shows the standard deviation of $\xi_\mathrm{21,LAE}(r)$ across the $8$ subboxes. Results are shown for varied global \HI fractions at $z=8.0$, $7.3$, $7.0$, $6.7$ from dark to bright colours. For these redshifts the \HI fractions are $\langle \chi_\mathrm{HI} \rangle=0.84$, $0.69$, $0.52$, $0.23$ for {\sc mhinc} and $0.71$, $0.59$, $0.49$, $0.34$ for the {\sc mhdec} simulations, respectively (see Tab. \ref{tab:astraeus}).}
    \label{fig:21cmLAEcrosscorrfunc_resolution_astraeus}
\end{figure*}

\section{Impact of limited volumes}
\label{sec:impact_of_limited_volumes}

With the analytic limits for the 21cm-LAE cross-correlation function $\xi_\mathrm{21,LAE}$ at hand, we can investigate the impact of the simulation box size on the $\xi_\mathrm{21,LAE}(r)$ values derived from the gridded simulation boxes via the cross-power spectra during different stages of reionisation. We note that while cosmic variance affects the 21cm-LAE cross-correlation amplitudes, our analytic estimates (Eq. \ref{eq:xi21LAE_analytic}) should remain valid for the corresponding $\langle\chi_\mathrm{HI} \rangle$ and $\langle 1+ \delta \rangle_\mathrm{HI}$ values in the selected volume. For this reason, in this section, convergence refers to the deviation of the via the cross-power spectra derived $\xi_\mathrm{21,LAE}(r)$ values from our analytic estimates. Therefore, 
in Fig. \ref{fig:21cmLAEcrosscorrfunc_resolution_astraeus} we show the $\xi_\mathrm{21,LAE}(r)$ for different simulation box sizes, ranging from $80h^{-1}$Mpc to $320h^{-1}$Mpc for the {\sc astraeus} simulations. We note that since the {\sc astraeus} simulations have only been run on a periodic $160h^{-1}$Mpc box, we derive the results for a $320h^{-1}$Mpc box by concatenating the $160h^{-1}$ box at the same redshift in all three directions. While this will not recover the large-scale power missed due to cosmic variance (impacting the large-scale reionisation topology), it provides a rough estimate of the large scale modes required to derive converged $\xi_\mathrm{21,LAE}(r)$ values.

To obtain $\xi_\mathrm{21,LAE}(r)$ for the $80h^{-1}$Mpc box, we divide the $160h^{-1}$Mpc box into $8$ subboxes. For each of the subboxes, we compute $\xi_\mathrm{21,LAE}(r)$. We show their mean value (dotted lines) and standard deviation (shaded areas) across the subboxes in Fig. \ref{fig:21cmLAEcrosscorrfunc_resolution_astraeus}. 
Here the standard deviation measures the cosmic variance of such a volume, while the deviation of the mean value to our analytic estimate for the 21cm-LAE cross-correlation functions (essentially represented by $\xi_\mathrm{21,LAE}(r)$ for the $320h^{-1}$Mpc box in Fig. \ref{fig:21cmLAEcrosscorrfunc_resolution_astraeus}) estimates the convergence. We briefly digress to discuss the effects of cosmic variance.
Listing average ionisation fraction $\langle\chi_\mathrm{HII}\rangle$ and the overdensity of neutral regions $\langle1+\delta\rangle_\mathrm{HI}$ for each subbox, we can see from Tables \ref{tab:global_values_mhdec_astraeus} and \ref{tab:global_values_mhinc_astraeus} in Appendix \ref{app_subvolumes} that the ionisation history has not converged in a volume of a $80h^{-1}$Mpc box: $\langle\chi_\mathrm{HII}\rangle$ varies around $\sim 2-7\%$, with the variation amplitude rising as reionisation progresses. As the ionisation fronts propagate from dense to less dense regions, simulation boxes with lower ionisation levels show higher average densities of the neutral regions. As expected, we find these values to predict by how much $\xi_\mathrm{21,LAE}(r)$ of a subbox exceeds or subceeds the main value across all subboxes, with the deviation being proportional to $\langle\chi_\mathrm{HI}\rangle \langle1+\delta\rangle_\mathrm{HI}|_\mathrm{subbox} - \langle\chi_\mathrm{HI}\rangle \langle1+\delta\rangle_\mathrm{HI}$, as expected from our analytic estimate (Eqn. \ref{eq:xir0}).

Next we discuss the convergence by comparing the $\xi_\mathrm{21,LAE}(r)$ results across the different simulation volumes. In Fig. \ref{fig:21cmLAEcrosscorrfunc_resolution_astraeus} we see that the strength of the anti-correlation between the 21cm signal and LAEs drops the more large-scale cross-power is missed due to a decreasing box size (going from solid to dashed to dotted lines). How much large-scale power is missed depends on (1) the global ionisation fraction $\langle\chi_\mathrm{HI}\rangle$ and (2) the reionisation topology, i.e. the correlation between the reionisation redshift $z_\mathrm{reion}$ and the underlying gas density field:

{\bf Dependence on $\langle \chi_\mathrm{HI} \rangle$:} For both {\sc astraeus} simulations the difference between the $160h^{-1}$cMpc or $80h^{-1}$cMpc to the $320h^{-1}$cMpc box increases as the Universe becomes more ionised and $\langle \chi_\mathrm{HI}\rangle$ decreases. It shifts from $\lesssim1\%$ ($\lesssim3\%$) at $z=8$ to $3-4\%$ ($5-6\%$) at $z=6.6$ for the $160h^{-1}$cMpc ($80h^{-1}$cMpc) box. The growth of the ionised regions in size enhances the importance of large-scale power. The closer their sizes become to those of the simulation box, the less volume is left to map the background accurately.

{\bf Dependence on $1+\delta$-$z_\mathrm{reion}$ cross-correlation:} The {\sc mhinc} scenario where brighter galaxies are the main drivers of reionisation shows larger differences in the small-scale 21cm-LAE cross-correlation amplitude among different box sizes than the {\sc mhdec} scenario (c.f. values at $\langle\chi_\mathrm{HI}\rangle\simeq0.5$ in {\sc mhdec} (orange lines) and {\sc mhinc} (medium blue lines)). The reason is similar to that for the dependence on $\langle\chi_\mathrm{HI}\rangle$: in the {\sc mhinc} scenario, the spatial variance of the ionising emissivity is higher as more massive galaxies have higher $f_\mathrm{esc}$ values, therefore the ionised regions around LAEs are larger and their sizes get closer to that of the simulation box at higher $\langle\chi_\mathrm{HI}\rangle$ values. 

Finally, for full convergence, i.e. the $\xi_\mathrm{21,LAE}(r)$ values derived via the cross-power spectra numerically agree with our analytical estimates, we find the {\sc astraeus} simulation box to be at the limit. Ideally, larger simulation box of $\sim 300h^{-1}$Mpc on the side would be required to obtain converged results. Interestingly, these volumes are in good agreement with those found necessary for the 21cm power spectrum to converge due to cosmic variance in previous studies \citep{Iliev2014, Kaur2020}. However, it should be also noted that the combined SKA-Subaru HSC observational uncertainties depicted in Fig. \ref{fig:crossr0} are of similar order than the deviation of the $80h^{-1}$cMpc box $\xi_\mathrm{21,LAE}(r)$ values from the $320h^{-1}$cMpc box results. Finally, we note that computing the 21cm-LAE cross-correlation functions not via the cross-power spectra but directly in real-space may be an avenue to avoid the convergence issues for smaller volumes, as obtaining the small-scale amplitudes does not rely then on capturing the large-scale fluctuations. 
We also find them more stable when deriving the 21cm-LAE cross-correlation functions from mock realisations that include the thermal noise in the 21cm signal maps. However, computing the 21cm-LAE cross-correlation functions in real-space is significantly slower than via the cross-power spectra (approximately hours versus minutes for the $1024^3$ grids of the EOS simulations) but should remain feasible for similar sized grids and probably more robust when cross-correlating future noisy 21cm maps with volume-limited LAE data.

\section{Comparison to previous work}
\label{sec:comparison_to_previous_work}

As we have seen in previous Sections, the 21cm-LAE cross-correlation functions are not only sensitive to the overall ionisation state of the IGM but also to the reionisation topology and the simulation box size. In the following, we compare our results to previous works and highlight why their predictions agree or differ from our analytic model. 

Firstly, all works where the field of the 21cm signal fluctuations, $\delta_{21}$, scales with $\delta T_b$ depict the $\langle\chi_\mathrm{HI}\rangle$-dependency of $\xi_\mathrm{21,LAE}(r)$ \citep[see][]{Vrbanec2016, Sobacchi2016, Hutter2017, Hutter2018b, Heneka2020, Vrbanec2020, Weinberger2020}. This scaling is not seen in \citet{Kubota2018} as their 21cm signal fluctuation field is normalised by $\langle\delta T_b\rangle$ and thus not sensitive to $\langle\chi_\mathrm{HI}\rangle \langle1+\delta\rangle_\mathrm{HI}$.
Although $\xi_\mathrm{21,LAE}(r)$ in \citet{Weinberger2020} is shown in units of mK, dividing their $\lim_{r \to 0}\xi_\mathrm{21,LAE}(r)$ values by $T_0$ (as given by their Eqn. 2) yields anti-correlation amplitudes that are in rough agreement with our analytic limits, with $\langle1+\delta\rangle_\mathrm{HI}$ decreasing from $\lesssim1$ at $z\simeq7.4$ to $\sim0.8$ at $z\simeq6.6$ for their Very Late model. Similarly, the 21cm-LAE cross-correlation results in \citet{Heneka2020} confirm our analytic limits for both the post-heating as well as heating epoch when $T_s\sim T_\mathrm{CMB}$. The results in \citet{Vrbanec2016} and \citet{Vrbanec2020} are also in line with our predictions, however dividing their $\xi_\mathrm{21,LAE}$ values at the smallest scales shown yields a constant $\langle1+\delta\rangle_\mathrm{HI}$ value during the second half of reionisation ($\langle\chi_\mathrm{HI}\rangle\lesssim0.5$).
Secondly, the 21cm-LAE cross-correlation predictions in some works \citep{Hutter2017, Hutter2018b, Kubota2018} show lower anti-correlation amplitudes due to their box sizes of $\lesssim200$cMpc and deriving the cross-correlations from the cross-power spectra. As a result, the survey parameters predicted in these works to distinguish between different stages of reionisation represent conservative limits.
Thirdly, \citet{Sobacchi2016} finds also lower anti-correlation amplitudes despite a sufficient volume of $\gtrsim500^3$cMpc$^3$. This might be due to actually showing the 2D 21cm-LAE cross-correlation functions or their method of connecting the intrinsic Ly$\alpha$ luminosity to halos or their LAEs being located in partially neutral regions (as their LAE model might allow sufficient IGM transmission of Lyman-$\alpha$ because of the redshifted Ly$\alpha$ line emerging from galaxies). The latter is likely to be also the main reason for the weaker 21cm-galaxy anti-correlation amplitudes in \citet{Park2014}. Their galaxy sample extends down to halo masses of $M_h\simeq2\times10^8\msun$, which are most abundant but not able to ionise a cell of $\sim0.5$cMpc length alone.

\section{Conclusions}
\label{sec:conclusions}

We have computed the 21cm-LAE cross-correlation functions, $\xi_\mathrm{21,LAE}(r)$, for different reionisation scenarios simulated with two different semi-numerical schemes following galaxy evolution and reionisation. While {\sc astraeus} derives the galaxy properties from the simulated DM mass assembly histories like semi-analytical galaxy evolution models, {\sc 21cmfast} follows a more semi-empirical approach to infer galaxy properties. The scenarios differ in the large-scale distribution of the ionising emissivity and cover the physically plausible range of (1) the escape fraction of ionising photons $f_\mathrm{esc}$ decreasing to increasing with halo mass and (2) different galaxy mass (virial temperature) thresholds for star formation. This diverse data set has allowed us to verify the analytic limit of the small-scale 21cm-LAE cross-correlation amplitude we derived and to propose a physically motivated analytic fitting function for the 21cm-LAE cross-correlation function during the EoR. Our fitting function assumes the sizes of ionised regions around LAEs to follow a lognormal distribution and fits the numerical results derived from the different simulations well. The analytic limit and fitting function for the 21cm-LAE cross-correlation function allow us to draw the following conclusions:
\begin{enumerate}
    \item The small-scale 21cm-LAE cross-correlation amplitude, $\xi_\mathrm{21,LAE}(r\simeq0)$, is directly proportional to the mean neutral hydrogen fraction and the average spin-temperature weighted overdensity in neutral regions. In the post-heating regime ($T_\mathrm{s}\gg T_\mathrm{CMB}$) the dependence on the spin temperature becomes negligible.
    \item Assuming a lognormal distribution for the sizes of the ionised regions provides a good approximation for the ionised regions around LAEs and allows us to analytically derive the shape of $\xi_\mathrm{21,LAE}(r)$ across all scales. The peak of the size distribution of the ionised regions and thus typical size of ionised regions around LAEs corresponds to the inversion point in $\xi_\mathrm{21,LAE}(r)$.
    \item Scaling with the average overdensity in neutral regions, the 21cm-LAE cross-correlation amplitude is also sensitive to the reionisation topology, i.e. the propagation of the ionisation fronts within the cosmic web structure. The stronger the emerging ionising emissivity is correlated to the underlying gas distribution, i.e. the more the ionising emissivity is biased to low-mass galaxies, the weaker is the 21cm-LAE anti-correlation amplitude.
    \item The smaller the simulation box is, the more large-scale modes are not contributing to the large-scale anti-correlation between the 21cm signal and LAEs, relevant when the 21cm-LAE cross-correlation function is derived via the 21cm-LAE cross-power spectrum and leading to the 21cm-LAE anti-correlation amplitude being underestimated. This effect increases with the size of the ionised regions and their size and distribution being sensitive to cosmic variance. We find that $\sim300h^{-1}$cMpc boxes provide large enough volumes for the numerically derived 21cm-LAE cross-correlation functions to reproduce our small-scale analytic limit.
    \item Our analytic predictions and volume studies can explain the different 21cm-LAE (21cm-galaxy) cross-correlation predictions to date. Given the information provided in previous works, we find them due to different normalisations of the 21cm fluctuation field $\delta_{21}$, too small simulation volumes, or galaxies/LAEs being located in partially neutral simulation cells (due to a too large cell size for the given galaxy population).
\end{enumerate}

The functional form of the 21cm-LAE cross-correlation function that we derived in this paper provides not only a test for future cross-correlation predictions from simulations, e.g. whether the simulated volume is sufficient, but also a fitting function for the 21cm-LAE cross-correlation functions derived from future observations. The latter could provide a quick way to derive constraints for reionisation. However, the problem remains that the 21cm-LAE anti-correlation amplitude is sensitive to both the ionisation history and topology. Breaking this degeneracy would require at least a tight relation between the size distribution of the ionised regions and the average ionisation level of the IGM. Future work, however, needs to show whether such a relation is sufficient or whether the non-Gaussian nature of the ionised large-scale structure needs to be accounted for. We already see that the size distributions of ionised regions do not differ very strongly for opposing $f_\mathrm{esc}$ scenarios. For this reason, 21cm LAE cross-correlations are unlikely to provide tighter constraints on reionisation unless they are complemented by analyses that trace the non-Gaussianity of the 21cm signal, such as the bispectrum \citep{Majumdar2018, Hutter2020, Majumdar2020, Tiwari2022} and other shape-sensitive statistics \citep{Gazagnes2021}.

Despite these shortcomings, our analytic representation of the 21cm-LAE cross-correlation function offers a computationally cheap way to predict which combinations and designs of 21cm and LAE surveys with forthcoming telescopes (e.g. SKA, Roman, Subaru's Hyper Suprime-Cam) would provide the best constrained 21cm-LAE cross-correlation functions and thus constraints on reionisation.

\section*{Acknowledgements}

AH acknowledges support by the VILLUM FONDEN under grant 37459. The Cosmic Dawn Center (DAWN) is funded by the Danish National Research Foundation under grant No. 140. CH acknowledges funding by Volkswagen Foundation and is supported by the Deutsche Forschungsgemeinschaft (DFG, German Research Foundation) under Germany's Excellence Strategy EXC 2181/1 -- 390900948 (the Heidelberg STRUCTURES Excellence Cluster). PD and MT acknowledge support from the NWO grant 016.VIDI.189.162 (``ODIN"). PD warmly thanks the European Commission's and University of Groningen's CO-FUND Rosalind Franklin program. 
GY acknowledges  Ministerio de  Ciencia e Innovaci\'on (Spain) for partial financial support under research grant PID2021-122603NB-C21. The {\sc vsmdpl} simulation has been performed at LRZ Munich within the project {\it pr87yi}. The CosmoSim data base (\url{www.cosmosim.org}) provides access to the simulation and the Rockstar data. The data base is a service by the Leibniz Institute for Astrophysics Potsdam (AIP).
This research made use of \texttt{matplotlib}, a Python library for publication quality graphics \citep{hunter2007}; and the Python library \texttt{numpy} \citep{numpy}.

\section*{Data Availability}

The {\sc astraeus} code is publicly available on GitHub (\url{https://github.com/annehutter/astraeus}, \citealt{Hutter2020astraeus}) as well as the EOS simulations on \url{http://homepage.sns.it/mesinger/EOS.html}. The full {\sc astraeus} and EOS simulations and the data derived from these data sets will be shared on reasonable request to the corresponding authors.



\bibliographystyle{mnras}
\bibliography{hutter_papers,papers} 

\begin{thebibliography}{}
\makeatletter
\relax
\def\mn@urlcharsother{\let\do\@makeother \do\$\do\&\do\#\do\^\do\_\do\%\do\~}
\def\mn@doi{\begingroup\mn@urlcharsother \@ifnextchar [ {\mn@doi@}
  {\mn@doi@[]}}
\def\mn@doi@[#1]#2{\def\@tempa{#1}\ifx\@tempa\@empty \href
  {http://dx.doi.org/#2} {doi:#2}\else \href {http://dx.doi.org/#2} {#1}\fi
  \endgroup}
\def\mn@eprint#1#2{\mn@eprint@#1:#2::\@nil}
\def\mn@eprint@arXiv#1{\href {http://arxiv.org/abs/#1} {{\tt arXiv:#1}}}
\def\mn@eprint@dblp#1{\href {http://dblp.uni-trier.de/rec/bibtex/#1.xml}
  {dblp:#1}}
\def\mn@eprint@#1:#2:#3:#4\@nil{\def\@tempa {#1}\def\@tempb {#2}\def\@tempc
  {#3}\ifx \@tempc \@empty \let \@tempc \@tempb \let \@tempb \@tempa \fi \ifx
  \@tempb \@empty \def\@tempb {arXiv}\fi \@ifundefined
  {mn@eprint@\@tempb}{\@tempb:\@tempc}{\expandafter \expandafter \csname
  mn@eprint@\@tempb\endcsname \expandafter{\@tempc}}}

\bibitem[\protect\citeauthoryear{{Barry}, {Hazelton}, {Sullivan}, {Morales}  \&
  {Pober}}{{Barry} et~al.}{2016}]{Barry2016}
{Barry} N.,  {Hazelton} B.,  {Sullivan} I.,  {Morales} M.~F.,   {Pober} J.~C.,
  2016, \mn@doi [\mnras] {10.1093/mnras/stw1380}, \href
  {https://ui.adsabs.harvard.edu/abs/2016MNRAS.461.3135B} {461, 3135}

\bibitem[\protect\citeauthoryear{{Barry} et~al.,}{{Barry}
  et~al.}{2019}]{Barry2019}
{Barry} N.,  et~al., 2019, \mn@doi [\apj] {10.3847/1538-4357/ab40a8}, \href
  {https://ui.adsabs.harvard.edu/abs/2019ApJ...884....1B} {884, 1}

\bibitem[\protect\citeauthoryear{{Beane}, {Villaescusa-Navarro}  \&
  {Lidz}}{{Beane} et~al.}{2019}]{Beane2019}
{Beane} A.,  {Villaescusa-Navarro} F.,   {Lidz} A.,  2019, \mn@doi [\apj]
  {10.3847/1538-4357/ab0a08}, \href
  {https://ui.adsabs.harvard.edu/abs/2019ApJ...874..133B} {874, 133}

\bibitem[\protect\citeauthoryear{{Behroozi}, {Wechsler}  \& {Wu}}{{Behroozi}
  et~al.}{2013a}]{behroozi2013_rs}
{Behroozi} P.~S.,  {Wechsler} R.~H.,   {Wu} H.-Y.,  2013a, \mn@doi [\apj]
  {10.1088/0004-637X/762/2/109}, \href
  {https://ui.adsabs.harvard.edu/abs/2013ApJ...762..109B} {762, 109}

\bibitem[\protect\citeauthoryear{{Behroozi}, {Wechsler}, {Wu}, {Busha},
  {Klypin}  \& {Primack}}{{Behroozi} et~al.}{2013b}]{behroozi2013_trees}
{Behroozi} P.~S.,  {Wechsler} R.~H.,  {Wu} H.-Y.,  {Busha} M.~T.,  {Klypin}
  A.~A.,   {Primack} J.~R.,  2013b, \mn@doi [\apj]
  {10.1088/0004-637X/763/1/18}, \href
  {https://ui.adsabs.harvard.edu/abs/2013ApJ...763...18B} {763, 18}

\bibitem[\protect\citeauthoryear{{Bosman} et~al.,}{{Bosman}
  et~al.}{2021}]{Bosman2021}
{Bosman} S. E.~I.,  et~al., 2021, arXiv e-prints, \href
  {https://ui.adsabs.harvard.edu/abs/2021arXiv210803699B} {p. arXiv:2108.03699}

\bibitem[\protect\citeauthoryear{{Carilli} \& {Rawlings}}{{Carilli} \&
  {Rawlings}}{2004}]{Carilli2004}
{Carilli} C.~L.,  {Rawlings} S.,  2004, \mn@doi [\nar]
  {10.1016/j.newar.2004.09.001}, \href
  {https://ui.adsabs.harvard.edu/abs/2004NewAR..48..979C} {48, 979}

\bibitem[\protect\citeauthoryear{{Castellano} et~al.,}{{Castellano}
  et~al.}{2016}]{Castellano2016}
{Castellano} M.,  et~al., 2016, \mn@doi [\apjl] {10.3847/2041-8205/818/1/L3},
  \href {https://ui.adsabs.harvard.edu/abs/2016ApJ...818L...3C} {818, L3}

\bibitem[\protect\citeauthoryear{{Castellano} et~al.,}{{Castellano}
  et~al.}{2018}]{Castellano2018}
{Castellano} M.,  et~al., 2018, \mn@doi [\apjl] {10.3847/2041-8213/aad59b},
  \href {https://ui.adsabs.harvard.edu/abs/2018ApJ...863L...3C} {863, L3}

\bibitem[\protect\citeauthoryear{{Croft} et~al.}{{Croft}
  et~al.}{2016}]{Croft16}
{Croft} R. A.~C.,  et~al., 2016, \mn@doi [\mnras] {10.1093/mnras/stw204}, \href
  {https://ui.adsabs.harvard.edu/abs/2016MNRAS.457.3541C} {457, 3541}

\bibitem[\protect\citeauthoryear{{Dayal} \& {Ferrara}}{{Dayal} \&
  {Ferrara}}{2018}]{Dayal2018}
{Dayal} P.,  {Ferrara} A.,  2018, \mn@doi [\physrep]
  {10.1016/j.physrep.2018.10.002}, \href
  {https://ui.adsabs.harvard.edu/abs/2018PhR...780....1D} {780, 1}

\bibitem[\protect\citeauthoryear{{Dayal}, {Maselli}  \& {Ferrara}}{{Dayal}
  et~al.}{2011}]{Dayal2011}
{Dayal} P.,  {Maselli} A.,   {Ferrara} A.,  2011, \mn@doi [\mnras]
  {10.1111/j.1365-2966.2010.17482.x}, \href
  {https://ui.adsabs.harvard.edu/abs/2011MNRAS.410..830D} {410, 830}

\bibitem[\protect\citeauthoryear{{Dayal}, {Ferrara}, {Dunlop}  \&
  {Pacucci}}{{Dayal} et~al.}{2014}]{Dayal2014}
{Dayal} P.,  {Ferrara} A.,  {Dunlop} J.~S.,   {Pacucci} F.,  2014, \mn@doi
  [\mnras] {10.1093/mnras/stu1848}, \href
  {https://ui.adsabs.harvard.edu/abs/2014MNRAS.445.2545D} {445, 2545}

\bibitem[\protect\citeauthoryear{{Dayal} et~al.,}{{Dayal}
  et~al.}{2022}]{Dayal2022}
{Dayal} P.,  et~al., 2022, \mn@doi [\mnras] {10.1093/mnras/stac537}, \href
  {https://ui.adsabs.harvard.edu/abs/2022MNRAS.512..989D} {512, 989}

\bibitem[\protect\citeauthoryear{{DeBoer} et~al.,}{{DeBoer}
  et~al.}{2017}]{DeBoer2017}
{DeBoer} D.~R.,  et~al., 2017, \mn@doi [\pasp]
  {10.1088/1538-3873/129/974/045001}, \href
  {https://ui.adsabs.harvard.edu/abs/2017PASP..129d5001D} {129, 045001}

\bibitem[\protect\citeauthoryear{{Dijkstra}, {Wyithe}, {Haiman}, {Mesinger}  \&
  {Pentericci}}{{Dijkstra} et~al.}{2014}]{Dijkstra2014}
{Dijkstra} M.,  {Wyithe} S.,  {Haiman} Z.,  {Mesinger} A.,   {Pentericci} L.,
  2014, \mn@doi [\mnras] {10.1093/mnras/stu531}, \href
  {https://ui.adsabs.harvard.edu/abs/2014MNRAS.440.3309D} {440, 3309}

\bibitem[\protect\citeauthoryear{{Dijkstra}, {Gronke}  \&
  {Venkatesan}}{{Dijkstra} et~al.}{2016}]{Dijkstra2016}
{Dijkstra} M.,  {Gronke} M.,   {Venkatesan} A.,  2016, \mn@doi [\apj]
  {10.3847/0004-637X/828/2/71}, \href
  {https://ui.adsabs.harvard.edu/abs/2016ApJ...828...71D} {828, 71}

\bibitem[\protect\citeauthoryear{{Field}}{{Field}}{1958}]{Field1958}
{Field} G.~B.,  1958, \mn@doi [Proceedings of the IRE]
  {10.1109/JRPROC.1958.286741}, \href
  {https://ui.adsabs.harvard.edu/abs/1958PIRE...46..240F} {46, 240}

\bibitem[\protect\citeauthoryear{{Fuller} et~al.,}{{Fuller}
  et~al.}{2020}]{Fuller2020}
{Fuller} S.,  et~al., 2020, \mn@doi [\apj] {10.3847/1538-4357/ab959f}, \href
  {https://ui.adsabs.harvard.edu/abs/2020ApJ...896..156F} {896, 156}

\bibitem[\protect\citeauthoryear{{Furlanetto} \& {Lidz}}{{Furlanetto} \&
  {Lidz}}{2007}]{Furlanetto_Lidz2007}
{Furlanetto} S.~R.,  {Lidz} A.,  2007, \mn@doi [\apj] {10.1086/513009}, \href
  {https://ui.adsabs.harvard.edu/abs/2007ApJ...660.1030F} {660, 1030}

\bibitem[\protect\citeauthoryear{{Furlanetto}, {Oh}  \& {Briggs}}{{Furlanetto}
  et~al.}{2006}]{Furlanetto2006}
{Furlanetto} S.~R.,  {Oh} S.~P.,   {Briggs} F.~H.,  2006, \mn@doi [\physrep]
  {10.1016/j.physrep.2006.08.002}, \href
  {https://ui.adsabs.harvard.edu/abs/2006PhR...433..181F} {433, 181}

\bibitem[\protect\citeauthoryear{{Gazagnes}, {Koopmans}  \&
  {Wilkinson}}{{Gazagnes} et~al.}{2021}]{Gazagnes2021}
{Gazagnes} S.,  {Koopmans} L. V.~E.,   {Wilkinson} M. H.~F.,  2021, \mn@doi
  [\mnras] {10.1093/mnras/stab107}, \href
  {https://ui.adsabs.harvard.edu/abs/2021MNRAS.502.1816G} {502, 1816}

\bibitem[\protect\citeauthoryear{{Goto} et~al.,}{{Goto}
  et~al.}{2021}]{Goto2021}
{Goto} H.,  et~al., 2021, \mn@doi [\apj] {10.3847/1538-4357/ac308b}, \href
  {https://ui.adsabs.harvard.edu/abs/2021ApJ...923..229G} {923, 229}

\bibitem[\protect\citeauthoryear{{Gronke}, {Dijkstra}, {McCourt}  \&
  {Oh}}{{Gronke} et~al.}{2017}]{Gronke2017}
{Gronke} M.,  {Dijkstra} M.,  {McCourt} M.,   {Oh} S.~P.,  2017, \mn@doi [\aap]
  {10.1051/0004-6361/201731013}, \href
  {https://ui.adsabs.harvard.edu/abs/2017A&A...607A..71G} {607, A71}

\bibitem[\protect\citeauthoryear{{Heneka} \& {Mesinger}}{{Heneka} \&
  {Mesinger}}{2020}]{Heneka2020}
{Heneka} C.,  {Mesinger} A.,  2020, \mn@doi [\mnras] {10.1093/mnras/staa1517},
  \href {https://ui.adsabs.harvard.edu/abs/2020MNRAS.496..581H} {496, 581}

\bibitem[\protect\citeauthoryear{{Heneka}, {Cooray}  \& {Feng}}{{Heneka}
  et~al.}{2017}]{Heneka2017}
{Heneka} C.,  {Cooray} A.,   {Feng} C.,  2017, \mn@doi [\apj]
  {10.3847/1538-4357/aa8eed}, \href
  {https://ui.adsabs.harvard.edu/abs/2017ApJ...848...52H} {848, 52}

\bibitem[\protect\citeauthoryear{Hunter}{Hunter}{2007}]{hunter2007}
Hunter J.~D.,  2007, \mn@doi [Computing In Science \& Engineering]
  {10.1109/MCSE.2007.55}, 9, 90

\bibitem[\protect\citeauthoryear{{Hutter}}{{Hutter}}{2018}]{Hutter2018a}
{Hutter} A.,  2018, \mn@doi [\mnras] {10.1093/mnras/sty683}, \href
  {https://ui.adsabs.harvard.edu/abs/2018MNRAS.477.1549H} {477, 1549}

\bibitem[\protect\citeauthoryear{{Hutter}, {Dayal}, {Partl}  \&
  {M{\"u}ller}}{{Hutter} et~al.}{2014}]{Hutter2014}
{Hutter} A.,  {Dayal} P.,  {Partl} A.~M.,   {M{\"u}ller} V.,  2014, \mn@doi
  [\mnras] {10.1093/mnras/stu791}, \href
  {https://ui.adsabs.harvard.edu/abs/2014MNRAS.441.2861H} {441, 2861}

\bibitem[\protect\citeauthoryear{{Hutter}, {Dayal}, {M{\"u}ller}  \&
  {Trott}}{{Hutter} et~al.}{2017}]{Hutter2017}
{Hutter} A.,  {Dayal} P.,  {M{\"u}ller} V.,   {Trott} C.~M.,  2017, \mn@doi
  [\apj] {10.3847/1538-4357/836/2/176}, \href
  {https://ui.adsabs.harvard.edu/abs/2017ApJ...836..176H} {836, 176}

\bibitem[\protect\citeauthoryear{{Hutter}, {Trott}  \& {Dayal}}{{Hutter}
  et~al.}{2018}]{Hutter2018b}
{Hutter} A.,  {Trott} C.~M.,   {Dayal} P.,  2018, \mn@doi [\mnras]
  {10.1093/mnrasl/sly115}, \href
  {https://ui.adsabs.harvard.edu/abs/2018MNRAS.479L.129H} {479, L129}

\bibitem[\protect\citeauthoryear{{Hutter}, {Watkinson}, {Seiler}, {Dayal},
  {Sinha}  \& {Croton}}{{Hutter} et~al.}{2020}]{Hutter2020}
{Hutter} A.,  {Watkinson} C.~A.,  {Seiler} J.,  {Dayal} P.,  {Sinha} M.,
  {Croton} D.~J.,  2020, \mn@doi [\mnras] {10.1093/mnras/stz3139}, \href
  {https://ui.adsabs.harvard.edu/abs/2020MNRAS.492..653H} {492, 653}

\bibitem[\protect\citeauthoryear{{Hutter}, {Dayal}, {Yepes}, {Gottl{\"o}ber},
  {Legrand}  \& {Ucci}}{{Hutter} et~al.}{2021}]{Hutter2021a}
{Hutter} A.,  {Dayal} P.,  {Yepes} G.,  {Gottl{\"o}ber} S.,  {Legrand} L.,
  {Ucci} G.,  2021, \mn@doi [\mnras] {10.1093/mnras/stab602}, \href
  {https://ui.adsabs.harvard.edu/abs/2021MNRAS.503.3698H} {503, 3698}

\bibitem[\protect\citeauthoryear{{Hutter}, {Trebitsch}, {Dayal},
  {Gottl{\"o}ber}, {Yepes}  \& {Legrand}}{{Hutter} et~al.}{2022}]{Hutter2022}
{Hutter} A.,  {Trebitsch} M.,  {Dayal} P.,  {Gottl{\"o}ber} S.,  {Yepes} G.,
  {Legrand} L.,  2022, arXiv e-prints, \href
  {https://ui.adsabs.harvard.edu/abs/2022arXiv220914592H} {p. arXiv:2209.14592}

\bibitem[\protect\citeauthoryear{{Iliev}, {Mellema}, {Ahn}, {Shapiro}, {Mao}
  \& {Pen}}{{Iliev} et~al.}{2014}]{Iliev2014}
{Iliev} I.~T.,  {Mellema} G.,  {Ahn} K.,  {Shapiro} P.~R.,  {Mao} Y.,   {Pen}
  U.-L.,  2014, \mn@doi [\mnras] {10.1093/mnras/stt2497}, \href
  {https://ui.adsabs.harvard.edu/abs/2014MNRAS.439..725I} {439, 725}

\bibitem[\protect\citeauthoryear{{Jensen}, {Laursen}, {Mellema}, {Iliev},
  {Sommer-Larsen}  \& {Shapiro}}{{Jensen} et~al.}{2013}]{Jensen2013}
{Jensen} H.,  {Laursen} P.,  {Mellema} G.,  {Iliev} I.~T.,  {Sommer-Larsen} J.,
    {Shapiro} P.~R.,  2013, \mn@doi [\mnras] {10.1093/mnras/sts116}, \href
  {https://ui.adsabs.harvard.edu/abs/2013MNRAS.428.1366J} {428, 1366}

\bibitem[\protect\citeauthoryear{{Kakiichi} \& {Gronke}}{{Kakiichi} \&
  {Gronke}}{2021}]{Kakiichi2021}
{Kakiichi} K.,  {Gronke} M.,  2021, \mn@doi [\apj] {10.3847/1538-4357/abc2d9},
  \href {https://ui.adsabs.harvard.edu/abs/2021ApJ...908...30K} {908, 30}

\bibitem[\protect\citeauthoryear{{Kaur}, {Gillet}  \& {Mesinger}}{{Kaur}
  et~al.}{2020}]{Kaur2020}
{Kaur} H.~D.,  {Gillet} N.,   {Mesinger} A.,  2020, \mn@doi [\mnras]
  {10.1093/mnras/staa1323}, \href
  {https://ui.adsabs.harvard.edu/abs/2020MNRAS.495.2354K} {495, 2354}

\bibitem[\protect\citeauthoryear{{Keating}, {Weinberger}, {Kulkarni},
  {Haehnelt}, {Chardin}  \& {Aubert}}{{Keating} et~al.}{2020}]{Keating2020}
{Keating} L.~C.,  {Weinberger} L.~H.,  {Kulkarni} G.,  {Haehnelt} M.~G.,
  {Chardin} J.,   {Aubert} D.,  2020, \mn@doi [\mnras] {10.1093/mnras/stz3083},
  \href {https://ui.adsabs.harvard.edu/abs/2020MNRAS.491.1736K} {491, 1736}

\bibitem[\protect\citeauthoryear{{Kimm}, {Blaizot}, {Garel}, {Michel-Dansac},
  {Katz}, {Rosdahl}, {Verhamme}  \& {Haehnelt}}{{Kimm} et~al.}{2019}]{Kimm2019}
{Kimm} T.,  {Blaizot} J.,  {Garel} T.,  {Michel-Dansac} L.,  {Katz} H.,
  {Rosdahl} J.,  {Verhamme} A.,   {Haehnelt} M.,  2019, \mn@doi [\mnras]
  {10.1093/mnras/stz989}, \href
  {https://ui.adsabs.harvard.edu/abs/2019MNRAS.486.2215K} {486, 2215}

\bibitem[\protect\citeauthoryear{{Klypin}, {Yepes}, {Gottl{\"o}ber}, {Prada}
  \& {He{\ss}}}{{Klypin} et~al.}{2016}]{Klypin2016}
{Klypin} A.,  {Yepes} G.,  {Gottl{\"o}ber} S.,  {Prada} F.,   {He{\ss}} S.,
  2016, \mn@doi [\mnras] {10.1093/mnras/stw248}, \href
  {https://ui.adsabs.harvard.edu/abs/2016MNRAS.457.4340K} {457, 4340}

\bibitem[\protect\citeauthoryear{{Kubota}, {Yoshiura}, {Takahashi}, {Hasegawa},
  {Yajima}, {Ouchi}, {Pindor}  \& {Webster}}{{Kubota}
  et~al.}{2018}]{Kubota2018}
{Kubota} K.,  {Yoshiura} S.,  {Takahashi} K.,  {Hasegawa} K.,  {Yajima} H.,
  {Ouchi} M.,  {Pindor} B.,   {Webster} R.~L.,  2018, \mn@doi [\mnras]
  {10.1093/mnras/sty1471}, \href
  {https://ui.adsabs.harvard.edu/abs/2018MNRAS.479.2754K} {479, 2754}

\bibitem[\protect\citeauthoryear{{Li} et~al.,}{{Li} et~al.}{2019}]{Li2019}
{Li} W.,  et~al., 2019, \mn@doi [\apj] {10.3847/1538-4357/ab55e4}, \href
  {https://ui.adsabs.harvard.edu/abs/2019ApJ...887..141L} {887, 141}

\bibitem[\protect\citeauthoryear{{Lidz}, {Zahn}, {Furlanetto}, {McQuinn},
  {Hernquist}  \& {Zaldarriaga}}{{Lidz} et~al.}{2009}]{Lidz2009}
{Lidz} A.,  {Zahn} O.,  {Furlanetto} S.~R.,  {McQuinn} M.,  {Hernquist} L.,
  {Zaldarriaga} M.,  2009, \mn@doi [\apj] {10.1088/0004-637X/690/1/252}, \href
  {https://ui.adsabs.harvard.edu/abs/2009ApJ...690..252L} {690, 252}

\bibitem[\protect\citeauthoryear{{Liu} \& {Shaw}}{{Liu} \&
  {Shaw}}{2020}]{Liu2020}
{Liu} A.,  {Shaw} J.~R.,  2020, \mn@doi [\pasp] {10.1088/1538-3873/ab5bfd},
  \href {https://ui.adsabs.harvard.edu/abs/2020PASP..132f2001L} {132, 062001}

\bibitem[\protect\citeauthoryear{{Maity} \& {Choudhury}}{{Maity} \&
  {Choudhury}}{2022}]{Maity2022}
{Maity} B.,  {Choudhury} T.~R.,  2022, \mn@doi [\mnras]
  {10.1093/mnras/stac1847}, \href
  {https://ui.adsabs.harvard.edu/abs/2022MNRAS.515..617M} {515, 617}

\bibitem[\protect\citeauthoryear{{Majumdar}, {Pritchard}, {Mondal},
  {Watkinson}, {Bharadwaj}  \& {Mellema}}{{Majumdar}
  et~al.}{2018}]{Majumdar2018}
{Majumdar} S.,  {Pritchard} J.~R.,  {Mondal} R.,  {Watkinson} C.~A.,
  {Bharadwaj} S.,   {Mellema} G.,  2018, \mn@doi [\mnras]
  {10.1093/mnras/sty535}, \href
  {https://ui.adsabs.harvard.edu/abs/2018MNRAS.476.4007M} {476, 4007}

\bibitem[\protect\citeauthoryear{{Majumdar}, {Kamran}, {Pritchard}, {Mondal},
  {Mazumdar}, {Bharadwaj}  \& {Mellema}}{{Majumdar}
  et~al.}{2020}]{Majumdar2020}
{Majumdar} S.,  {Kamran} M.,  {Pritchard} J.~R.,  {Mondal} R.,  {Mazumdar} A.,
  {Bharadwaj} S.,   {Mellema} G.,  2020, \mn@doi [\mnras]
  {10.1093/mnras/staa3168}, \href
  {https://ui.adsabs.harvard.edu/abs/2020MNRAS.499.5090M} {499, 5090}

\bibitem[\protect\citeauthoryear{{Mauerhofer} \& {Dayal}}{{Mauerhofer} \&
  {Dayal}}{2023}]{Mauerhofer2023}
{Mauerhofer} V.,  {Dayal} P.,  2023, \mn@doi [arXiv e-prints]
  {10.48550/arXiv.2305.01681}, \href
  {https://ui.adsabs.harvard.edu/abs/2023arXiv230501681M} {p. arXiv:2305.01681}

\bibitem[\protect\citeauthoryear{{McQuinn}, {Lidz}, {Zahn}, {Dutta},
  {Hernquist}  \& {Zaldarriaga}}{{McQuinn} et~al.}{2007}]{McQuinn2007}
{McQuinn} M.,  {Lidz} A.,  {Zahn} O.,  {Dutta} S.,  {Hernquist} L.,
  {Zaldarriaga} M.,  2007, \mn@doi [\mnras] {10.1111/j.1365-2966.2007.11489.x},
  \href {https://ui.adsabs.harvard.edu/abs/2007MNRAS.377.1043M} {377, 1043}

\bibitem[\protect\citeauthoryear{{Meerburg}, {Dvorkin}  \&
  {Spergel}}{{Meerburg} et~al.}{2013}]{Meerburg2013}
{Meerburg} P.~D.,  {Dvorkin} C.,   {Spergel} D.~N.,  2013, \mn@doi [\apj]
  {10.1088/0004-637X/779/2/124}, \href
  {https://ui.adsabs.harvard.edu/abs/2013ApJ...779..124M} {779, 124}

\bibitem[\protect\citeauthoryear{{Mertens}, {Ghosh}  \& {Koopmans}}{{Mertens}
  et~al.}{2018}]{Mertens2018}
{Mertens} F.~G.,  {Ghosh} A.,   {Koopmans} L.~V.~E.,  2018, \mn@doi [\mnras]
  {10.1093/mnras/sty1207}, \href
  {https://ui.adsabs.harvard.edu/abs/2018MNRAS.478.3640M} {478, 3640}

\bibitem[\protect\citeauthoryear{{Mertens} et~al.,}{{Mertens}
  et~al.}{2020}]{Mertens2020}
{Mertens} F.~G.,  et~al., 2020, \mn@doi [\mnras] {10.1093/mnras/staa327}, \href
  {https://ui.adsabs.harvard.edu/abs/2020MNRAS.493.1662M} {493, 1662}

\bibitem[\protect\citeauthoryear{{Mesinger} \& {Furlanetto}}{{Mesinger} \&
  {Furlanetto}}{2008}]{Mesinger2008}
{Mesinger} A.,  {Furlanetto} S.~R.,  2008, \mn@doi [\mnras]
  {10.1111/j.1365-2966.2008.13039.x}, \href
  {https://ui.adsabs.harvard.edu/abs/2008MNRAS.386.1990M} {386, 1990}

\bibitem[\protect\citeauthoryear{{Mesinger}, {Aykutalp}, {Vanzella},
  {Pentericci}, {Ferrara}  \& {Dijkstra}}{{Mesinger}
  et~al.}{2015}]{Mesinger2015}
{Mesinger} A.,  {Aykutalp} A.,  {Vanzella} E.,  {Pentericci} L.,  {Ferrara} A.,
    {Dijkstra} M.,  2015, \mn@doi [\mnras] {10.1093/mnras/stu2089}, \href
  {https://ui.adsabs.harvard.edu/abs/2015MNRAS.446..566M} {446, 566}

\bibitem[\protect\citeauthoryear{{Mesinger}, {Greig}  \& {Sobacchi}}{{Mesinger}
  et~al.}{2016}]{Mesinger2016}
{Mesinger} A.,  {Greig} B.,   {Sobacchi} E.,  2016, \mn@doi [\mnras]
  {10.1093/mnras/stw831}, \href
  {https://ui.adsabs.harvard.edu/abs/2016MNRAS.459.2342M} {459, 2342}

\bibitem[\protect\citeauthoryear{{Mineo}, {Gilfanov}  \& {Sunyaev}}{{Mineo}
  et~al.}{2012}]{Mineo12}
{Mineo} S.,  {Gilfanov} M.,   {Sunyaev} R.,  2012, \mn@doi [\mnras]
  {10.1111/j.1365-2966.2012.21831.x}, \href
  {http://adsabs.harvard.edu/abs/2012MNRAS.426.1870M} {426, 1870}

\bibitem[\protect\citeauthoryear{Oliphant}{Oliphant}{2006}]{numpy}
Oliphant T.,  2006, {NumPy}: A guide to {NumPy}, USA: Trelgol Publishing, \url
  {http://www.numpy.org/}

\bibitem[\protect\citeauthoryear{{Ouchi} et~al.,}{{Ouchi}
  et~al.}{2010}]{Ouchi2010}
{Ouchi} M.,  et~al., 2010, \mn@doi [\apj] {10.1088/0004-637X/723/1/869}, \href
  {https://ui.adsabs.harvard.edu/abs/2010ApJ...723..869O} {723, 869}

\bibitem[\protect\citeauthoryear{{Ouchi} et~al.,}{{Ouchi}
  et~al.}{2018}]{Ouchi2018}
{Ouchi} M.,  et~al., 2018, \mn@doi [\pasj] {10.1093/pasj/psx074}, \href
  {https://ui.adsabs.harvard.edu/abs/2018PASJ...70S..13O} {70, S13}

\bibitem[\protect\citeauthoryear{{Park}, {Kim}, {Wyithe}  \& {Lacey}}{{Park}
  et~al.}{2014}]{Park2014}
{Park} J.,  {Kim} H.-S.,  {Wyithe} J. S.~B.,   {Lacey} C.~G.,  2014, \mn@doi
  [\mnras] {10.1093/mnras/stt2366}, \href
  {https://ui.adsabs.harvard.edu/abs/2014MNRAS.438.2474P} {438, 2474}

\bibitem[\protect\citeauthoryear{{Patil} et~al.,}{{Patil}
  et~al.}{2016}]{Patil2016}
{Patil} A.~H.,  et~al., 2016, \mn@doi [\mnras] {10.1093/mnras/stw2277}, \href
  {https://ui.adsabs.harvard.edu/abs/2016MNRAS.463.4317P} {463, 4317}

\bibitem[\protect\citeauthoryear{{Patil} et~al.,}{{Patil}
  et~al.}{2017}]{Patil2017}
{Patil} A.~H.,  et~al., 2017, \mn@doi [\apj] {10.3847/1538-4357/aa63e7}, \href
  {https://ui.adsabs.harvard.edu/abs/2017ApJ...838...65P} {838, 65}

\bibitem[\protect\citeauthoryear{{Pentericci} et~al.,}{{Pentericci}
  et~al.}{2014}]{Pentericci2014}
{Pentericci} L.,  et~al., 2014, \mn@doi [\apj] {10.1088/0004-637X/793/2/113},
  \href {https://ui.adsabs.harvard.edu/abs/2014ApJ...793..113P} {793, 113}

\bibitem[\protect\citeauthoryear{{Pentericci} et~al.,}{{Pentericci}
  et~al.}{2018}]{Pentericci2018}
{Pentericci} L.,  et~al., 2018, \mn@doi [\aap] {10.1051/0004-6361/201732465},
  \href {https://ui.adsabs.harvard.edu/abs/2018A&A...619A.147P} {619, A147}

\bibitem[\protect\citeauthoryear{{Planck Collaboration} et~al.,}{{Planck
  Collaboration} et~al.}{2016}]{Planck2016}
{Planck Collaboration} et~al., 2016, \mn@doi [\aap]
  {10.1051/0004-6361/201525830}, \href
  {https://ui.adsabs.harvard.edu/abs/2016A&A...594A..13P} {594, A13}

\bibitem[\protect\citeauthoryear{{Planck Collaboration} et~al.,}{{Planck
  Collaboration} et~al.}{2020}]{Planck2020}
{Planck Collaboration} et~al., 2020, \mn@doi [\aap]
  {10.1051/0004-6361/201833910}, \href
  {https://ui.adsabs.harvard.edu/abs/2020A&A...641A...6P} {641, A6}

\bibitem[\protect\citeauthoryear{{Pober} et~al.}{{Pober}
  et~al.}{2013}]{Pober13}
{Pober} J.~C.,  et~al., 2013, \mn@doi [\aj] {10.1088/0004-6256/145/3/65}, \href
  {https://ui.adsabs.harvard.edu/\#abs/2013AJ....145...65P} {145, 65}

\bibitem[\protect\citeauthoryear{{Pober} et~al.}{{Pober}
  et~al.}{2014}]{Pober14}
{Pober} J.~C.,  et~al., 2014, \mn@doi [\apj] {10.1088/0004-637X/782/2/66},
  \href {http://adsabs.harvard.edu/abs/2014ApJ...782...66P} {782, 66}

\bibitem[\protect\citeauthoryear{{Qin}, {Mesinger}, {Bosman}  \& {Viel}}{{Qin}
  et~al.}{2021}]{Qin2021}
{Qin} Y.,  {Mesinger} A.,  {Bosman} S. E.~I.,   {Viel} M.,  2021, \mn@doi
  [\mnras] {10.1093/mnras/stab1833}, \href
  {https://ui.adsabs.harvard.edu/abs/2021MNRAS.506.2390Q} {506, 2390}

\bibitem[\protect\citeauthoryear{{Qin}, {Wyithe}, {Oesch}, {Illingworth},
  {Leonova}, {Mutch}  \& {Naidu}}{{Qin} et~al.}{2022}]{Qin2022}
{Qin} Y.,  {Wyithe} J. S.~B.,  {Oesch} P.~A.,  {Illingworth} G.~D.,  {Leonova}
  E.,  {Mutch} S.~J.,   {Naidu} R.~P.,  2022, \mn@doi [\mnras]
  {10.1093/mnras/stab3733}, \href
  {https://ui.adsabs.harvard.edu/abs/2022MNRAS.510.3858Q} {510, 3858}

\bibitem[\protect\citeauthoryear{{Schenker}, {Ellis}, {Konidaris}  \&
  {Stark}}{{Schenker} et~al.}{2014}]{Schenker2014}
{Schenker} M.~A.,  {Ellis} R.~S.,  {Konidaris} N.~P.,   {Stark} D.~P.,  2014,
  \mn@doi [\apj] {10.1088/0004-637X/795/1/20}, \href
  {https://ui.adsabs.harvard.edu/abs/2014ApJ...795...20S} {795, 20}

\bibitem[\protect\citeauthoryear{{Shaver}, {Windhorst}, {Madau}  \& {de
  Bruyn}}{{Shaver} et~al.}{1999}]{Shaver1999}
{Shaver} P.~A.,  {Windhorst} R.~A.,  {Madau} P.,   {de Bruyn} A.~G.,  1999,
  \aap, \href {https://ui.adsabs.harvard.edu/abs/1999A&A...345..380S} {345,
  380}

\bibitem[\protect\citeauthoryear{{Sobacchi} \& {Mesinger}}{{Sobacchi} \&
  {Mesinger}}{2014}]{sobacchi14}
{Sobacchi} E.,  {Mesinger} A.,  2014, \mn@doi [\mnras] {10.1093/mnras/stu377},
  \href {http://adsabs.harvard.edu/abs/2014MNRAS.440.1662S} {440, 1662}

\bibitem[\protect\citeauthoryear{{Sobacchi}, {Mesinger}  \& {Greig}}{{Sobacchi}
  et~al.}{2016}]{Sobacchi2016}
{Sobacchi} E.,  {Mesinger} A.,   {Greig} B.,  2016, \mn@doi [\mnras]
  {10.1093/mnras/stw811}, \href
  {https://ui.adsabs.harvard.edu/abs/2016MNRAS.459.2741S} {459, 2741}

\bibitem[\protect\citeauthoryear{{Springel}}{{Springel}}{2005}]{Springel2005}
{Springel} V.,  2005, \mn@doi [\mnras] {10.1111/j.1365-2966.2005.09655.x},
  \href {http://adsabs.harvard.edu/abs/2005MNRAS.364.1105S} {364, 1105}

\bibitem[\protect\citeauthoryear{{Tiwari}, {Shaw}, {Majumdar}, {Kamran}  \&
  {Choudhury}}{{Tiwari} et~al.}{2022}]{Tiwari2022}
{Tiwari} H.,  {Shaw} A.~K.,  {Majumdar} S.,  {Kamran} M.,   {Choudhury} M.,
  2022, \mn@doi [\jcap] {10.1088/1475-7516/2022/04/045}, \href
  {https://ui.adsabs.harvard.edu/abs/2022JCAP...04..045T} {2022, 045}

\bibitem[\protect\citeauthoryear{{Trott} \& {Wayth}}{{Trott} \&
  {Wayth}}{2016}]{Trott2016}
{Trott} C.~M.,  {Wayth} R.~B.,  2016, \mn@doi [\pasa] {10.1017/pasa.2016.18},
  \href {https://ui.adsabs.harvard.edu/abs/2016PASA...33...19T} {33, e019}

\bibitem[\protect\citeauthoryear{{Ucci} et~al.,}{{Ucci}
  et~al.}{2021}]{Ucci2022}
{Ucci} G.,  et~al., 2021, arXiv e-prints, \href
  {https://ui.adsabs.harvard.edu/abs/2021arXiv211202115U} {p. arXiv:2112.02115}

\bibitem[\protect\citeauthoryear{{Verhamme}, {Orlitov{\'a}}, {Schaerer}  \&
  {Hayes}}{{Verhamme} et~al.}{2015}]{Verhamme2015}
{Verhamme} A.,  {Orlitov{\'a}} I.,  {Schaerer} D.,   {Hayes} M.,  2015, \mn@doi
  [\aap] {10.1051/0004-6361/201423978}, \href
  {https://ui.adsabs.harvard.edu/abs/2015A&A...578A...7V} {578, A7}

\bibitem[\protect\citeauthoryear{{Vrbanec} et~al.,}{{Vrbanec}
  et~al.}{2016}]{Vrbanec2016}
{Vrbanec} D.,  et~al., 2016, \mn@doi [\mnras] {10.1093/mnras/stv2993}, \href
  {https://ui.adsabs.harvard.edu/abs/2016MNRAS.457..666V} {457, 666}

\bibitem[\protect\citeauthoryear{{Vrbanec}, {Ciardi}, {Jeli{\'c}}, {Jensen},
  {Iliev}, {Mellema}  \& {Zaroubi}}{{Vrbanec} et~al.}{2020}]{Vrbanec2020}
{Vrbanec} D.,  {Ciardi} B.,  {Jeli{\'c}} V.,  {Jensen} H.,  {Iliev} I.~T.,
  {Mellema} G.,   {Zaroubi} S.,  2020, \mn@doi [\mnras]
  {10.1093/mnras/staa183}, \href
  {https://ui.adsabs.harvard.edu/abs/2020MNRAS.492.4952V} {492, 4952}

\bibitem[\protect\citeauthoryear{{Weinberger}, {Kulkarni}  \&
  {Haehnelt}}{{Weinberger} et~al.}{2020}]{Weinberger2020}
{Weinberger} L.~H.,  {Kulkarni} G.,   {Haehnelt} M.~G.,  2020, \mn@doi [\mnras]
  {10.1093/mnras/staa749}, \href
  {https://ui.adsabs.harvard.edu/abs/2020MNRAS.494..703W} {494, 703}

\bibitem[\protect\citeauthoryear{{Wiersma} et~al.,}{{Wiersma}
  et~al.}{2013}]{Wiersma2013}
{Wiersma} R.~P.~C.,  et~al., 2013, \mn@doi [\mnras] {10.1093/mnras/stt624},
  \href {https://ui.adsabs.harvard.edu/abs/2013MNRAS.432.2615W} {432, 2615}

\bibitem[\protect\citeauthoryear{{Wouthuysen}}{{Wouthuysen}}{1952}]{Wouthuysen1952}
{Wouthuysen} S.~A.,  1952, \mn@doi [\aj] {10.1086/106661}, \href
  {https://ui.adsabs.harvard.edu/abs/1952AJ.....57R..31W} {57, 31}

\bibitem[\protect\citeauthoryear{{Wyithe}, {Loeb}  \& {Schmidt}}{{Wyithe}
  et~al.}{2007}]{Wyithe2007}
{Wyithe} J. S.~B.,  {Loeb} A.,   {Schmidt} B.~P.,  2007, \mn@doi [\mnras]
  {10.1111/j.1365-2966.2007.12149.x}, \href
  {https://ui.adsabs.harvard.edu/abs/2007MNRAS.380.1087W} {380, 1087}

\bibitem[\protect\citeauthoryear{{Zahn}, {Lidz}, {McQuinn}, {Dutta},
  {Hernquist}, {Zaldarriaga}  \& {Furlanetto}}{{Zahn} et~al.}{2007}]{Zahn2007}
{Zahn} O.,  {Lidz} A.,  {McQuinn} M.,  {Dutta} S.,  {Hernquist} L.,
  {Zaldarriaga} M.,   {Furlanetto} S.~R.,  2007, \mn@doi [\apj]
  {10.1086/509597}, \href
  {https://ui.adsabs.harvard.edu/abs/2007ApJ...654...12Z} {654, 12}

\bibitem[\protect\citeauthoryear{{Zhu} et~al.,}{{Zhu} et~al.}{2021}]{Zhu2021}
{Zhu} Y.,  et~al., 2021, \mn@doi [\apj] {10.3847/1538-4357/ac26c2}, \href
  {https://ui.adsabs.harvard.edu/abs/2021ApJ...923..223Z} {923, 223}

\makeatother
\end{thebibliography}



\appendix

\section{Cross-correlation amplitude for subvolumes}
\label{app_subvolumes}

In \ref{tab:global_values_mhdec_astraeus} and \ref{tab:global_values_mhinc_astraeus} we list the average ionisation fraction $\langle\chi_\mathrm{HII}\rangle$ and neutral overdensity in each of the $8$ $80h^{-1}$cMpc subboxes at redshifts $z=8.0$, $7.3$, $7.0$ and $6.7$ for the {\sc mhdec} and {\sc mhinc} scenarios, respectively. Vales exceeding the average values of the entire simulation box ($160h^{-1}$cMpc, second column) are marked in red, while those falling short are marked in blue. The more a subbox is ionised, the lower its average overdensity in neutral regions, leading to a lower 21cm-LAE cross-correlation amplitude. The resulting variance in the 21cm-LAE cross-correlation amplitude across the $8$ subboxes is shown as shaded regions in Fig. \ref{fig:21cmLAEcrosscorrfunc_resolution_astraeus}.

\begin{table*}
    \centering
    \begin{tabular}{c|c|c|c|c|c|c|c|c|c|c}
        \hline
        z & $\langle\chi_\mathrm{HII}\rangle$ & $\langle\chi_\mathrm{HII}\rangle^{0,0,0}$ & $\langle\chi_\mathrm{HII}\rangle^{1,0,0}$ & $\langle\chi_\mathrm{HII}\rangle^{0,1,0}$ &
        $\langle\chi_\mathrm{HII}\rangle^{0,0,1}$ & $\langle\chi_\mathrm{HII}\rangle^{1,1,0}$ & $\langle\chi_\mathrm{HII}\rangle^{1,0,1}$ & $\langle\chi_\mathrm{HII}\rangle^{0,1,1}$ & $\langle\chi_\mathrm{HII}\rangle^{1,1,1}$ & $\sigma_{\langle\chi_\mathrm{HII}\rangle}$ \\
        \hline
        8.0 & 0.290 & \textcolor{blue}{0.281} & \textcolor{red}{0.313} & \textcolor{blue}{0.268} & \textcolor{red}{0.308} & \textcolor{red}{0.302} & \textcolor{blue}{0.261} & \textcolor{blue}{0.287} & \textcolor{red}{0.296} & 0.019\\
        7.3 & 0.413 & \textcolor{blue}{0.400} & \textcolor{red}{0.446} & \textcolor{blue}{0.381} & \textcolor{red}{0.445} & \textcolor{red}{0.433} & \textcolor{blue}{0.371} & \textcolor{blue}{0.409} & \textcolor{red}{0.420} & 0.028 \\
        7.0 & 0.511 & \textcolor{blue}{0.497} & \textcolor{red}{0.548} & \textcolor{blue}{0.468} & \textcolor{red}{0.558} & \textcolor{red}{0.541} & \textcolor{blue}{0.455} & \textcolor{blue}{0.501} & \textcolor{red}{0.521} & 0.038 \\
        6.6 & 0.660 & \textcolor{blue}{0.653} & \textcolor{red}{0.715} & \textcolor{blue}{0.602} & \textcolor{red}{0.723} & \textcolor{red}{0.702} & \textcolor{blue}{0.577} & \textcolor{blue}{0.635} & \textcolor{red}{0.673} & 0.053\\
        \hline
        \hline
        z & $\langle1+\delta\rangle_\mathrm{HI}$ & $\langle1+\delta\rangle_\mathrm{HI}^{0,0,0}$ & $\langle1+\delta\rangle_\mathrm{HI}^{1,0,0}$ & $\langle1+\delta\rangle_\mathrm{HI}^{0,1,0}$ & $\langle1+\delta\rangle_\mathrm{HI}^{0,0,1}$ & $\langle1+\delta\rangle_\mathrm{HI}^{1,1,0}$ & $\langle1+\delta\rangle_\mathrm{HI}^{1,0,1}$ & $\langle1+\delta\rangle_\mathrm{HI}^{0,1,1}$ & $\langle1+\delta\rangle_\mathrm{HI}^{1,1,1}$ & $\sigma_{\langle1+\delta\rangle_\mathrm{HI}}$ \\
        \hline
        8.0 &  0.781 & \textcolor{red}{0.788} & \textcolor{blue}{0.773} & \textcolor{red}{0.795} & \textcolor{blue}{0.771} & \textcolor{blue}{0.774} & \textcolor{red}{0.793} & \textcolor{red}{0.783} & \textcolor{blue}{0.775} & 0.0096\\
        7.3 & 0.730 & \textcolor{red}{0.737} & \textcolor{blue}{0.721} & \textcolor{red}{0.746} & \textcolor{blue}{0.719} & \textcolor{blue}{0.722} & \textcolor{red}{0.744} & \textcolor{red}{0.732} & \textcolor{blue}{0.723} & 0.0108 \\
        7.0 & 0.698 & \textcolor{red}{0.705} & \textcolor{blue}{0.688} & \textcolor{red}{0.714} & \textcolor{blue}{0.686} & \textcolor{blue}{0.689} & \textcolor{red}{0.713} & \textcolor{red}{0.700} & \textcolor{blue}{0.690} & 0.0115\\
        6.6 & 0.661 & \textcolor{red}{0.669} & \textcolor{blue}{0.654} & \textcolor{red}{0.679} & \textcolor{blue}{0.649} & \textcolor{blue}{0.651} & \textcolor{red}{0.677} & \textcolor{red}{0.664} & \textcolor{blue}{0.652} & 0.0121 \\
        \hline
    \end{tabular}
    \caption{Global hydrogen ionisation fraction $\langle\chi_\mathrm{HII}\rangle$ and mean overdensity in neutral regions $\langle1+\delta\rangle_\mathrm{HI}$ at different redshifts in the {\sc mhdec} {\sc astraeus} simulation (column 2). Columns $3-10$ show the respective values for the $8$ subboxes with each having a length of $80h^{-1}$Mpc. Column $11$ depicts the standard deviation across the $8$ subboxes of the respective values.}
    \label{tab:global_values_mhdec_astraeus}
\end{table*}

\begin{table*}
    \centering
    \begin{tabular}{c|c|c|c|c|c|c|c|c|c|c}
        \hline
        z & $\langle\chi_\mathrm{HII}\rangle$ & $\langle\chi_\mathrm{HII}\rangle^{0,0,0}$ & $\langle\chi_\mathrm{HII}\rangle^{1,0,0}$ & $\langle\chi_\mathrm{HII}\rangle^{0,1,0}$ &
        $\langle\chi_\mathrm{HII}\rangle^{0,0,1}$ & $\langle\chi_\mathrm{HII}\rangle^{1,1,0}$ & $\langle\chi_\mathrm{HII}\rangle^{1,0,1}$ & $\langle\chi_\mathrm{HII}\rangle^{0,1,1}$ & $\langle\chi_\mathrm{HII}\rangle^{1,1,1}$ & $\sigma_{\langle\chi_\mathrm{HII}\rangle}$ \\
        \hline
        8.0 & 0.157 & \textcolor{blue}{0.156} & \textcolor{red}{0.179} & \textcolor{blue}{0.137} & \textcolor{red}{0.166} & \textcolor{red}{0.166} & \textcolor{blue}{0.140} & \textcolor{blue}{0.153} & \textcolor{red}{0.161} & 0.014 \\
        7.3 & 0.313 & \textcolor{blue}{0.313} & \textcolor{red}{0.359} & \textcolor{blue}{0.267} & \textcolor{red}{0.339} & \textcolor{red}{0.339} & \textcolor{blue}{0.278} & \textcolor{blue}{0.297} & \textcolor{red}{0.314} & 0.032 \\
        7.0 & 0.482 & \textcolor{blue}{0.497} & \textcolor{red}{0.545} & \textcolor{blue}{0.405} & \textcolor{red}{0.522} & \textcolor{red}{0.523} & \textcolor{blue}{0.427} & \textcolor{blue}{0.453} & \textcolor{red}{0.484} & 0.062 \\
        6.6 & 0.770 & \textcolor{blue}{0.804} & \textcolor{red}{0.839} & \textcolor{blue}{0.671} & \textcolor{red}{0.795} & \textcolor{red}{0.820} & \textcolor{blue}{0.686} & \textcolor{blue}{0.739} & \textcolor{red}{0.808} & 0.064\\
        \hline
        \hline
        z & $\langle1+\delta\rangle_\mathrm{HI}$ & $\langle1+\delta\rangle_\mathrm{HI}^{0,0,0}$ & $\langle1+\delta\rangle_\mathrm{HI}^{1,0,0}$ & $\langle1+\delta\rangle_\mathrm{HI}^{0,1,0}$ & $\langle1+\delta\rangle_\mathrm{HI}^{0,0,1}$ & $\langle1+\delta\rangle_\mathrm{HI}^{1,1,0}$ & $\langle1+\delta\rangle_\mathrm{HI}^{1,0,1}$ & $\langle1+\delta\rangle_\mathrm{HI}^{0,1,1}$ & $\langle1+\delta\rangle_\mathrm{HI}^{1,1,1}$ & $\sigma_{\langle1+\delta\rangle_\mathrm{HI}}$ \\
        \hline
        8.0 &  0.892 & \textcolor{red}{0.895} & \textcolor{blue}{0.885} & \textcolor{red}{0.902} & \textcolor{blue}{0.886} & \textcolor{blue}{0.887} & \textcolor{red}{0.900} & \textcolor{red}{0.894} & \textcolor{blue}{0.889} & 0.0065\\
        7.3 & 0.843 & \textcolor{red}{0.847} & \textcolor{blue}{0.834} & \textcolor{red}{0.856} & \textcolor{blue}{0.833} & \textcolor{blue}{0.835} & \textcolor{red}{0.852} & \textcolor{red}{0.847} & \textcolor{blue}{0.839} & 0.0081 \\
        7.0 & 0.806 & \textcolor{red}{0.810} & \textcolor{blue}{0.796} & \textcolor{red}{0.821} & \textcolor{blue}{0.794} & \textcolor{blue}{0.797} & \textcolor{red}{0.817} & \textcolor{red}{0.812} & \textcolor{blue}{0.802} & 0.0103\\
        6.6 & 0.757 & \textcolor{red}{0.757} & \textcolor{blue}{0.750} & \textcolor{red}{0.774} & \textcolor{blue}{0.746} & \textcolor{blue}{0.743} & \textcolor{red}{0.775} & \textcolor{red}{0.773} & \textcolor{blue}{0.740} & 0.0148 \\
        \hline
    \end{tabular}
    \caption{Global hydrogen ionisation fraction $\langle\chi_\mathrm{HII}\rangle$ and mean overdensity in neutral regions $\langle1+\delta\rangle_\mathrm{HI}$ at different redshifts in the {\sc mhinc} {\sc astraeus} simulation (column 2). Columns $3-10$ show the respective values for the $8$ subboxes with each having a length of $80h^{-1}$Mpc. Column $11$ depicts the standard deviation across the $8$ subboxes of the respective values.}
    \label{tab:global_values_mhinc_astraeus}
\end{table*}


\bsp	
\label{lastpage}
\end{document}